\documentclass[useAMS,usenatbib,referee]{biom}
\usepackage{amsmath}
\usepackage{amssymb}
\usepackage{booktabs}
\usepackage{multirow}
\usepackage{footnote}
\usepackage{float}
\usepackage{comment}
\usepackage{url}
\usepackage[figuresright]{rotating}
\usepackage{graphicx}
\usepackage{xcolor}
\usepackage[colorlinks=true,linkcolor=red,citecolor=blue]{hyperref}
\usepackage{subcaption}
\usepackage{caption}
\usepackage{xr}

\newcommand{\indep}{$\(\perp\!\!\!\perp\)$}
\newcommand{\ATE}{\mbox{\tiny{ATE}}}
\newcommand{\ATT}{\mbox{\tiny{ATT}}}
\newcommand{\ATO}{\mbox{\tiny{ATO}}}

\title[Causal Framework of External Controls]{Leveraging External Controls in Clinical Trials: Estimands, Estimation, Assumptions}

\author {Bo Liu$^{1,*}$\email{bo.liu1997@duke.edu},  
 Laine Thomas$^{2,**}$\email{laine.thomas@duke.edu}, Rury R. Holman$^{3,***}$\email{rury.holman@dtu.ox.ac.uk}, Fan Li$^{1,****}$\email{fl35@duke.edu} \\
$^{1}$ Department of Statistical Science, Duke University, Durham, North Carolina 27708, U.S.A.\\
$^{2}$ Department of Biostatistics and Bioinformatics, Duke University, Durham, North Carolina 27708, U.S.A \\
$^{3}$ Radcliffe Department of Medicine, Oxford University, Oxford, United Kingdom \\
}

\begin{document}

\date{{\it Received XXX} 20XX. {\it Revised XXX} 20XX.  {\it
Accepted XXX} 20XX.}



\pagerange{\pageref{firstpage}--\pageref{lastpage}} 
\volume{XX}
\pubyear{20XX}
\artmonth{XXX}


\doi{XXX}

\label{firstpage}

\begin{abstract}
It is increasingly common to augment randomized controlled trial with  external controls from observational data, to evaluate the treatment effect of an intervention. Traditional approaches to treatment effect estimation involve ambiguous estimands and unrealistic or strong assumptions, such as mean exchangeability. We introduce a double-indexed notation for potential outcomes to define causal estimands transparently and clarify distinct sources of implicit bias.  We show that the concurrent control arm is critical in assessing the plausibility of assumptions and providing unbiased causal estimation. We derive a consistent and locally efficient estimator for a class of weighted average treatment effect estimands that combines concurrent and external data without assuming mean exchangeability. This estimator incorporates an estimate of the systematic difference in outcomes between the concurrent and external units, of which we propose a Frish-Waugh-Lovell style partial regression method to obtain. We compare the proposed methods with existing methods using extensive simulation and applied to cardiovascular clinical trials.               
\end{abstract}

\begin{keywords}
causal inference, estimand, external control, efficient influence function, propensity score, weighting 
\end{keywords}

\maketitle

\section{Introduction}

Randomized controlled trials (RCT) are the gold standard for evaluating the efficacy and safety of new treatments. But RCTs have numerous well-known limitations. For example, it is often challenging to recruit patients, particularly with a placebo control arm in trials targeting new treatment for rare diseases or severe diseases such as late-stage cancer. Moreover, recruiting patients into RCTs might take a long time, which complicates analyzing and interpreting results in fast-evolving diseases. In the face of these challenges, there is increasing interest in study designs that augment RCTs with external controls (EC) from observational data, such as electronic health records, past trials, and diseases registries, where the subjects had no access to the treatment of interest \citep{baumfeld2020trial}. Proper utilization of ECs can provide evidence of effectiveness and safety and accelerate development of new treatments. However, external data are generally different from RCT data. A key first step is to curate and harmonize the different data sources to align the definition and scale of outcomes, treatments and covariates \citep[e.g.][]{burcu2020real}. Then the analyst must adopt proper statistical procedures to evaluate the treatment effects.

Recent work on RCT and EC integration increasingly aligns with the estimand framework outlined in ICH E9 guideline on statistical principles for clinical trials \citep{ICHE9}. Estimands are particularly important in this context because the combination of samples from different populations introduces a variety of relevant estimands, each with distinct generalizability. But some earlier methods, e.g., the Bayesian dynamic pooling methods, were generally opaque on the definition of estimands \citep[e.g.][]{hobbs2012commensurate,banbeta2019modified, wang2022propensity}.
A stream of recent research adopts the potential outcome framework in causal inference to elucidate the estimands and assumptions for integrating RCTs and EC data \citep[e.g.][]{cheng2023enhancing,li2023improving,gao2023integrating,colnet2024causal,guo2024adaptive}. These works commonly assume mean exchangeability  (ME) of the potential outcomes, namely, conditional on observed covariates, there is no difference in the potential outcomes in the absence of the treatment between the units in the concurrent and external study. However, ME is often implausible in practice, leading to biased treatment effect estimates. 

Moreover, as we shall show later, the conventional single-index notation of potential outcomes blurs treatment and the source of data (concurrent versus external). Single-index notation obscures the strength of the assumption of ME on which recent literature relies \citep{li2023improving,gao2023integrating,colnet2024causal}. In this paper, we first introduce a double-indexed notation for potential outcomes, based on which we define a new average treatment effect estimand (Section \ref{sec:exchange}). Similar notation has been proposed recently in the context of the generalizability of clinical trials \citep{dahabreh2019generalizing,ung2024generalizing}.  We clarify that the concurrent control arm is critical in assessing the plausibility of ME. We provide a weaker version of ME for identifying the causal estimands (Section \ref{sec:identification}) and further provide estimation strategies when ME is violated (Section \ref{sec:no-exchange}). In particular, we derive a consistent and locally efficient semiparametric estimator that combines concurrent and external data without assuming ME. This estimator requires an estimate of the systematic difference in the outcomes between the concurrent and external units, which we propose a novel two-step Frish-Waugh-Lovell type partial regression method to obtain. We further extend the method to a class of weighted average treatment effects estimands (Section \ref{sec:WATE}). Finally, we conduct simulations to examine the performance of the proposed method and compare it with existing methods (Section \ref{sec:simulation}), and apply the method to a cardiovascular clinical trial (Section \ref{sec:application}).

\section{Double-indexed potential outcomes and estimands} \label{sec:exchange}
Consider a clinical trial that targets the effect of a treatment on an outcome. Such a clinical trial (referred to as concurrent trial hereafter) may be designed to be: (i) a randomized controlled trial with an active treatment arm and a, possibly small, control arm; or (ii) a single-arm trial, where all units are assigned to the active treatment. Under both settings, it may be desirable to augment the small (or non-existing) control arm by external data. We will show it is critical to have a control arm in the concurrent trial, regardless of the sample size, to identify and estimate the treatment effect.

\subsection{Notation and Consistency}
Suppose we have two distinct samples and for each subject $i$, let $Z_i$ denote the data source: $Z_i = 1$ and $Z_i = 0$ indicate being in the concurrent and external sample, respectively.  If the former has $N_1$ patients and the latter $N_0$, the combined data has $N = N_1 + N_0$ total patients labeled by $i = 1, 2, \dots, N$.  Let $A_i$ denote the treatment status: $A_i = 1$ and $A_i = 0$ indicate receiving the active treatment and the control, respectively. In this context, the paired data ($Z$, $A$) have three possible combinations (1,1), (1,0), and (0,0) because external data are used only for controls and (0,1) can not occur. Under the single-arm trial setting, $A_i = Z_i$, so only (1,1) and (0,0) can occur.
For each unit, we observe a vector of pre-treatment covariates $X_i$, and an outcome $Y_i$.

We use potential outcomes to define causal estimands. The existing literature commonly adopts a single-indexed notation $Y_i(a)$ to denote the potential outcome under one of two treatment levels $a=0,1$, along with the assumption of consistency, $Y_i=Y_i(1)A_i+Y_i(0)(1-A_i)$  \citep{cole2009consistency, li2023improving, colnet2024causal}.  Consistency is a sub-assumption of the SUTVA \citep{Rubin80} and means that the observed outcome $Y_i$ equals $Y_i(a)$ when $A_i=a$ for $a=0,1$. 
However, this notation implicitly equates the outcome of a unit in the control arm of the concurrent trial with that in the external data. Specifically, the consistency assumption has a particularly strong interpretation because dependency of the potential outcomes on $Z_i$ is excluded. This is acceptable  only if we are willing to assume that the outcomes have been perfectly harmonized across $Z$.  

In order to define perfect harmonization and avoid implicit assumptions, we introduce a double-indexed notation, $Y^{z}_i(a)$, for potential outcomes, where $Y^{1}_i(a)$ is the potential outcome for subject $i$, as would be measured in the concurrent trial, and $Y^{0}_i(a)$ is the potential outcome for subject $i$, as would be measured in the external control, under fixed intervention level $a$.  This allows for the possibility that $Y^{1}_i(a)$ may not equal $Y^{0}_i(a)$. This could arise, for example, if the outcome were a biomarker and the contributing samples used different assays, or if the same clinical outcome were adjudicated differently across studies. More generally, differences in definition, implementation, or measurement accuracy could yield different versions of the potential outcomes across studies. Whether the difference is negligible or substantial will be an important consideration. We formally make an extended version of the consistency assumption:
\begin{assumption}[Composite consistency]\label{ass:consistency} For all $i$,
\begin{equation}
    Y_{i} = Y_i^{1}(1)A_iZ_i  + Y_i^{1}(0)(1-A_i)Z_i + Y_i^{0}(0)(1-A_i)(1-Z_i).
\end{equation}
\end{assumption}
Consequently, each subject $i$ has three potential outcomes $Y^1_i(1)$, $Y^1_i(0)$ and $Y^0_i(0)$, of which only the one corresponding to the observed data source and treatment status is observed as $Y_i$. This allows us to define perfect harmonization of the outcome across studies by $Y^1_i(0)=Y^0_i(0)=Y_i(0)$, in which case consistency simplifies to $Y_i=Y_i(1)A_i+Y_i(0)(1-A_i)$. Subsequently, we use the more general form of Assumption \ref{ass:consistency}.

\subsection{Estimands using the double-indexed potential outcomes}
Several common causal estimands characterize the effect of the treatment, differing in the target population.   An average treatment effect (ATE) is typically defined as the mean difference in potential outcomes with and without treatment among the population corresponding to the study sample. Extending the usual notation, we might define the ATE as:
\begin{equation} \label{eq:ATE}
    \tau^{\ATE} = \mathbb{E}[Y^1_i(1) - Y^1_i(0)].
\end{equation}
However, interpretation of this estimand is ambiguous. First, fixing $z=1$ within $Y^1_i(a)$ indicates that the outcome of interest corresponds to the concurrent trial. That aligns with viewing the trial as the gold standard for outcome assessment.  Second, the expectation is implicitly taken over a combined population of concurrent trial and external controls, as the individuals, $i$, arise from both
populations. The corresponding estimand does not align with either the concurrent trial population nor the external control population, but an opaque mixture of the two that depends on sample sizes. Therefore, the ATE is unappealing in this context.

A straightforward and interpretable alternative is to target the average treatment effect among the population represented in the concurrent trial by conditioning on $Z_i = 1$:   
\begin{equation} \label{eq:ATT}
    \tau^{\ATT} = \mathbb{E}[Y^{1}_i(1) - Y^{1}_i(0) \mid Z_i = 1].
\end{equation}
The estimand $\tau^{\ATT}$ fixes both the outcome and the population to match the concurrent trial. Note that superscript $z=1$ defines \textit{what version of the outcome} is targeted, whereas condition $Z_{i}=1$ defines \textit{what population} is targeted. This is akin to the average treatment effect among the treated (ATT) estimand in the standard causal inference literature. Generalization to other target populations will be discussed in Section \ref{sec:WATE}. Initially, we focus on $\tau^{\ATT}$ as the target estimand for both single-arm and two-arm concurrent trials.     

\subsection{Implicit estimands and bias}

Previously, estimands have been defined in single-indexed notation, including $\tau = \mathbb{E}[Y_i(1) - Y_i(0)]$ and $\tau_{1} = \mathbb{E}[Y_i(1) - Y_i(0) \mid Z_i = 1]$ \citep{li2023improving, colnet2024causal}.  These have several potential interpretations. Suppose $\tau_1$ is interpreted as  $\mathbb{E}[Y^1_i(1) - Y^0_i(0) \mid Z_i = 1]$. This is the estimand that is most straightforward to identify from single-arm concurrent trial data where only pairs $(Z,A) = \{(0,0),(1,1)\}$ are observed.  However, unless $Y^1_i(a) = Y^0_i(a) =  Y_i(a)$, it implies a contrast of different outcomes under treatment versus under control and is not a valid causal quantity.  

Moreover, $\tau_{1} $ can be decomposed into two parts: 
\begin{equation} \label{eq:decomposition}
   \tau_{1} = \mathbb{E}[Y^1_i(1) - Y^1_i(0) \mid Z_i = 1]+ \mathbb{E}[Y^1_i(0) - Y^0_i(0) \mid Z_i = 1]=\tau^{\ATT}+b
\end{equation}
where the first component is the $\tau^{\ATT}$ defined in \eqref{eq:ATT}, and the second component, $b$, is the mean difference in the potential outcomes measured in the concurrent vs external study, in the absence of the treatment, which we will refer to as the systematic difference between the external and concurrent outcome. Neither component can be identified without additional assumptions. This decomposition emphasizes that the single-indexed notation implicitly assumes away the systematic difference in potential outcomes, namely imposes $b=0$. 

\section{Identification of $\tau^{\ATT}$ and estimation with mean exchangability} \label{sec:identification}

\subsection{Identifying assumptions}

The following assumptions are standard in the literature of external control; we modify them to the double-index notation. 

\begin{assumption}[Unconfounded treatment assignment]\label{ass:unconfoundA} The active treatment is unconfounded in the concurrent trial conditional on covariates:
    $Y^1_i(a) \indep A_i \mid \{X_i, Z_i = 1\}$, or equivalently, $\Pr(Y^1_i(a)\mid X_i, Z_i=1, A_i=1)=\Pr(Y^1_i(a)\mid X_i, Z_i=1, A_i=0)$ for  $a=0,1$. 
\end{assumption} 

\begin{assumption}[Unconfounded trial participation]\label{ass:unconfoundZ} Whether a control subject participates in the concurrent trial or is represented in the external control is independent of their potential outcomes given the observed covariates: $Y^z_i(0) \indep Z_i \mid X_i$, or equivalently, $\Pr(Y^z_i(0)\mid Z_i=1, X_i)=\Pr(Y^z_i(0)\mid  Z_i=0, X_i)=\Pr(Y^z_i(0)\mid X_i)$ for $z=0,1$. 
\end{assumption}

Under Assumption \ref{ass:unconfoundA} and  \ref{ass:unconfoundZ}, the probability that each subject would be in the treatment arm and would participate in the concurrent trial may depend on the observed covariates. Consequently, we define two propensity scores, for being in the concurrent trial and the active treatment, respectively:
\begin{equation}
   e_{Zi}\equiv e_Z(X_i) = \Pr(Z_i = 1\mid X_i), \quad e_{Ai}\equiv  e_A(X_i) = \Pr(A_i = 1 \mid X_i, Z_i = 1).
\end{equation}

\begin{assumption}[Overlap]\label{ass:overlap} Each subject has a non-zero probability of being in the concurrent study, $e_{Zi} > 0$,  and a non-zero probability of being in the active treatment or control arm if the subject is in the concurrent trial, $0 < e_{Ai} < 1$ for all $i$.  
\end{assumption}

Consider a two-arm concurrent trial alone, where no external data i s used.  For a well-conducted study, the assumptions of consistency, unconfounded treatment assignment and overlap (Assumptions \ref{ass:consistency}, \ref{ass:unconfoundA} and \ref{ass:overlap}) hold by design. Accordingly, $\tau^{\ATT}$ can be non-parametrically identified by $\mathbb{E}(Y_i \mid Z_i = 1,A_i = 1) - \mathbb{E}(Y_i \mid Z_i = 1,A_i = 0)$. Non-parametric estimators as well as standard weighting, outcome modeling, and doubly-robust estimators can be constructed for $\tau^{\ATT}$ \citep{bang2005doubly}.

To augment the two-arm concurrent trial with external data, an additional assumption is necessary to relate $Y^0_i(0)$ to $Y^1_i(0)$.  Rather than assuming $Y^1_i(0) = Y^0_i(0)$, we specify mean exchangeability of the potential outcomes (ME-PO).   
\begin{assumption}[Mean exchangeability of the potential outcomes]\label{ass:mean-exchangeability} For all $z$ and $x$,
\begin{equation}
    b_z(x)\equiv \mathbb{E}[Y^1_i(0) \mid Z_i=z, X_i=x] - \mathbb{E}[Y^0_i(0) \mid  Z_i=z, X_i=x]=0. 
    \label{eq:ME}
\end{equation}
\end{assumption}  
Note that ME-PO pertains to the comparability, on average, of the potential outcomes, \textit{as measured} in the concurrent trial versus external data, for a fixed population with $Z_i = z$.  It excludes, on average, different versions of potential outcomes.   

\subsection{Connection to existing work} 

Previous literature has defined a slightly different version of ME \citep{li2023improving, colnet2024causal}:
\begin{assumption}[Mean exchangeability]\label{ass:cmean-exchangeability} For all $x$,
\begin{equation}
    b(x)\equiv \mathbb{E}[Y^{1}_i(0) \mid Z_i=1, X_i=x] - \mathbb{E}[Y^{0}_i(0) \mid Z_i=0, X_i=x]=0. 
    \label{eq:cME}
\end{equation}
\end{assumption}  
The difference between ME-PO and ME is that $b_z(x)$ compares different versions of the potential outcome for a fixed population ($Z_i=z$), whereas $b(x)$ contrasts two different populations ($Z_i=1$ vs $Z_i=0$). We can expand $b(x)$ into two separate components as follows:
\begin{align*}
      b(x) =   &  \{\mathbb{E}[Y^{0}_i(0) \mid Z_i=1, X_i=x] - \mathbb{E}[Y^{0}_i(0) \mid Z_i=0, X_i=x]\}
       + \\
     & \{ \mathbb{E}[Y^{1}_i(0) \mid Z_i=1, X_i=x] - \mathbb{E}[Y^{0}_i(0) \mid Z_i=1, X_i=x] \}. 
\end{align*} 
The first term is zero by definition of unconfounded trial participation (Assumption \ref{ass:unconfoundZ}). The second term equals $b_{1}(x)$ and is zero under ME-PO (Assumption \ref{ass:mean-exchangeability}). Thus ME is implied by ME-PO paired with Assumption \ref{ass:unconfoundZ}.  However, Assumption \ref{ass:unconfoundZ} is not always made explicit, and then ME is a stronger assumption than ME-PO. The deconstruction allows us to evaluate each assumption separately.  

\subsection{Estimation}

Below we use $\widehat{f}$ to generically denote a consistent estimate of a function $f$, e.g. $\widehat{\mu}_{zai}$ is a consistent estimate of the true outcome function $\mu_{zai}=\mathbb{E}[Y_i  \mid Z_i=z, A_i=a, X_i = x]$.  Under Assumptions \ref{ass:consistency} -- \ref{ass:mean-exchangeability}, \cite{li2023improving} and \cite{gao2023integrating} derived a locally efficient estimator for $\tau^{\ATT}$:
\begin{equation}\label{eq:ATT-estimator-aug}
    \widehat{\tau}^{\ATT}_{aug} = \frac{1}{N_1} \sum_{i=1}^N \left\{Z_i [\widehat{\mu}_{11i} - \widehat{\mu}_{10i}] + \frac{1}{\widehat{e}_{Ai}} \widehat{R}_{11i} - \dfrac{\widehat{e}_{Zi}}{1 - \widehat{e}_{Ai}\widehat{e}_{Zi}}[  \widehat{R}_{10i} +  \widehat{R}_{00i}] \right\},
\end{equation} 
where $\widehat{R}_{11i} = Z_iA_i[Y_i - \widehat{\mu}_{11i}]$, $\widehat{R}_{10i} = Z_i(1-A_i)[Y_i - \widehat{\mu}_{10i}]$, $\widehat{R}_{00i} = (1-Z_i)[Y_i - \widehat{\mu}_{00i}]$. Under ME, the values $\widehat{\mu}_{10i}$ and $\widehat{\mu}_{00i}$ are estimated jointly because $b(x) = 0$ implies that $\mu_{10i} = \mu_{00i}=\mu_{.0i}$. This increases the sample size and reduces the variance, but will introduce bias if ME does not hold. Estimator \eqref{eq:ATT-estimator-aug} is locally efficient when the variances are exchangeable in addition to the means, i.e. $\mathbb{V}[Y^1_i(0) \mid X_i] = \mathbb{V}[Y^0_i(0) \mid X_i]$. Exchangeability of the variance is supported by the design of externally controlled studies, and implied by ME for binary outcomes, although it can easily be relaxed. 

\section{Estimation without mean exchangeability} \label{sec:no-exchange}

The ME assumption is testable only if we have both concurrent and external control units. Under Assumptions \ref{ass:consistency}-\ref{ass:unconfoundA}, we can identify $b(X_{i})$ as
\begin{equation}\label{eq:bX}
    b(X_{i}) = \mathbb{E}(Y_{i} \mid Z_{i} = 1, A_{i} = 0, X_{i}) - \mathbb{E}(Y_{i} \mid Z_{i} = 0, A_{i} = 0, X_{i}),
\end{equation}
(see Supplementary Material for the proof). But $b(X_{i})$ is not testable in a single-arm trial and is thus easily subject to bias.  When $b(X_{i}) \neq 0$, \cite{li2023improving} argue that the bias will be bounded and will not affect the validity of $\widehat{\tau}^{\ATT}_{aug}$ when $b(X_{i})$ does not deviate far from 0. \cite{gao2023integrating} proposed to augment only a selective part of the external data where ME is deemed plausible. However, both proposals hinge on the ME assumption, even when the observed data may suggest otherwise. Formula \eqref{eq:bX} suggests that the observed data can not only reveal whether ME holds or not, but also identify the bias when it does not hold. This motivates us to estimate $b(X_{i})$ rather than simply test or assume $b(X_{i}) = 0$.

Note $b(X_{i})$ may be nonzero if either Assumption \ref{ass:unconfoundZ} or \ref{ass:mean-exchangeability} are violated.  So estimating this quantity permits departures from either assumption. If we are confident that unconfoundedness (Assumption \ref{ass:unconfoundZ}) holds, then we might interpret $b(X_{i})$ to reflect the difference in mean potential outcomes (Equation \ref{eq:ME}). Otherwise, $b(X_{i})$ captures both sources of bias.

\subsection{Known $b(X)$} 

We first illustrate how to use $b(X)$ when it is known, but not necessarily zero. The function $b(X) =\mu_{10}(X) - \mu_{00}(X)$ is not directly estimated in Equation \eqref{eq:ATT} for $\widehat{\tau}^{\ATT}_{aug}$. However, the known $b(X)$ is important to the estimation of $\mu_{10}(X)$, which otherwise might be unstable if the $(1, 0)$ group is small. With a regression outcome model $\mathbb{E}[Y^1(0) \mid X; \beta] = \mu_{10}(X; \beta)$, knowing $b(X)$ implies $\mathbb{E}[Y^0(0) \mid X; \beta] = \mu_{00}(X; \beta)= \mu_{10}(X; \beta) - b(X)$ and leads to the following estimating equation 
\begin{equation}\label{eq:estimating-equation-beta}
    \sum_{i=1}^N C(X_i)\{Z_i(1 - A_i)[Y_i - \mu_{10}(X_i; \beta)] + (1 - Z_i)[Y_i + b(X_i) - \mu_{10}(X_i;\beta)]\} = 0,
\end{equation}
          for an arbitrary function $C(x)$ (see Supplementary Material for proof). This estimating equation uses both external and concurrent controls to estimate $\beta$, and thus is generally more efficient than that using only the concurrent controls. Assume $\widehat{\beta}$ is the only solution to \eqref{eq:estimating-equation-beta} and converges in probability to $\beta$. Then we can plug in $\widehat{\mu}_{10i} = \mu_{10}(X_i; \widehat{\beta})$ and $\widehat{\mu}_{00i} = \mu_{10}(X_i; \widehat{\beta}) - b(X_i)$ into the estimator $\widehat{\tau}_{aug}^{\ATT}$ in \eqref{eq:ATT-estimator-aug} and $\widehat{\tau}_{aug}^{\ATT}$ remains consistent.

\subsection{Unknown $b(X)$} \label{sec:unknown-b}
In most real applications, $b(X)$ is unknown and needs to be estimated from the observed data. There are two extremes in estimating $b(X)$. One end is to estimate $\mu_{10}(X)$ and $\mu_{00}(X)$ separately, with the concurrent and external controls, respectively, and then obtain $\hat{b}(X)=\hat{\mu}_{10}(X) - \hat{\mu}_{00}(X)$. This procedure borrows no information from the external data in estimating $\mu_{10}(X)$. 
The other end is to fully specify $b(X)$, for example $b(X)=0$, and use all external and concurrent controls to estimate $\mu_{10}(X)$. This estimate will have smaller variance than that not using external controls, but it will be biased if $b(X)$ is misspecified.  

As an intermediate approach, we propose to impose a parametric model $b(X; \theta)$ for $b(X)$ whose complexity controls the degree of information borrowing. Determining the complexity amounts to a variance-bias trade-off. 
Here the model fit of $\mu_{10}(X)$ and $\mu_{00}(X)$ are not independent of each other due to the parametric form of $b$, and hence the external control units help estimate the model parameter $\beta$. Given a consistent estimate $\widehat{\theta}$ of $\theta$, we shall replace the unknown $b(X_i)$ in Equation \eqref{eq:estimating-equation-beta} by $b(X_i; \widehat{\theta})$ to further estimate $\mu_{10}(X; \beta)$. However, estimation of $b(x)$ can be challenging because of the small sample size of the concurrent control arm. Once $b(x)$ is estimated, $\mu_{10}(x)$ is easier to estimate because the external control units can be included. Therefore, it may also be tempting to assume a simple parametric model $b(x; \theta)$ for $b(x)$ while allowing a flexible specification for $\mu_{10}(x)$, e.g. semi- or non-parametric models. 
The proposed estimation procedure is summarized as follows.
\begin{enumerate}
    \item Fit a semi-parametric model with control units in the concurrent and external data to obtain a consistent estimate $\widehat{\theta}$ of $\theta$.
    \item For each unit in the concurrent control arm, \emph{i.e.}, $Z_i=1, A_i = 0$, obtain the pseudo-outcome as $\widetilde{Y}_i = Z_i(1-A_i)Y_i + (1 - Z_i)[Y_i - b(X_i; \widehat{\theta})]$.
    \item Fit a flexible mean regression model for the pseudo-outcomes $\widetilde{Y}_i$ of the concurrent control units, $\widehat{\mu}_{10}(X) = \mathbb{E}[\widetilde{Y} \mid X; \beta]$.
    \item Plug the estimated $\hat{b}(X)$ and $\widehat{\mu}_{10}(X)$ into the estimator $\widehat{\tau}^{\ATT}_{aug}$ in \eqref{eq:ATT-estimator-aug}.
\end{enumerate}

Step (2) and (3) are straightforward given $\widehat{\theta}$. For Step (1), many approaches are available for estimating $\widehat{\theta}$. Below we propose a two-step approach that works well in practice. Given equation \eqref{eq:bX}, we specify a model for $\mathbb{E}[Y \mid X, Z, A = 0]$. Our rationale is that, though the respective outcome function $\mathbb{E}[Y \mid X, Z=1, A = 0]$ and $\mathbb{E}[Y \mid X, Z=0, A = 0]$ for concurrent controls and for external controls may be complex, their difference $b(X)$ may be simple. Therefore, we specify the following model:
\begin{equation}\label{eq:cond-expectation}
    \mathbb{E}[Y \mid X, Z, A = 0] = \alpha(X) + b(X; \theta)Z,
\end{equation}
where $b(X; \theta)$ is a simple parametric model for the difference between $Z=1$ and
$Z=0$ group, while the term $\alpha(X)$ is an unspecified, potentially complex, non-parametric function for the baseline outcome. Common choices of $b(X;\theta)$ include $b(X;\theta) = \theta$ for constant difference or $b(X; \theta) = \theta_0 + X'\theta_1$ for linear difference in $X$. Specification of $b(X)$ shall depend on the proportion of the concurrent control arm: we may specify a more flexible model with a sizeable concurrent control arm because this is more data to stably estimate $b(x)$, but a simpler model otherwise.

Our interest lies in estimating the finite-dimensional parameter $\theta$.
Rewrite \eqref{eq:cond-expectation} as 
\begin{equation}\label{eq:YX}
    Y_i = \alpha(X_i) + b(X_i; \beta)Z_i + \varepsilon_i, \quad \mathbb{E}[\varepsilon_i \mid X_i, Z_i, A_i = 0] = 0
\end{equation}
for all $i$ with $A_i = 0$.
The conditional expectation on $X_i$ is 
\begin{equation} \label{eq:ExpYX}
    \mathbb{E}[Y_i \mid X_i, A_i = 0] = \alpha(X_i) + b(X_i;\theta) \mathbb{E}[Z_i \mid X_i, A_i = 0].
\end{equation}
Taking the difference between equation \eqref{eq:YX} and \eqref{eq:ExpYX}, the term $\alpha(X_i)$ cancels out, and we obtain
\begin{equation}
    Y_i - \mathbb{E}[Y_i \mid X_i, A_i = 0] = b(X_i; \theta)(Z_i - \mathbb{E}[Z_i\mid X_i, A_i = 0]) + \epsilon_i, \quad \mathbb{E}[\epsilon_i \mid X_i, Z_i, A_i = 0] = 0.
\end{equation}
This suggests a two-step partial regression in the Frisch-Waugh-Lovell \citep{yule1907theory} fashion: (1) estimate the conditional mean $\mathbb{E}[Y_i \mid X_i, A_i = 0]$ and $\mathbb{E}[Z_i \mid X_i, A_i = 0]$ and obtain the residuals, denoted by $\widehat{U}$ and $\widehat{V}$, respectively, using only the control units, and (2) estimate $\theta$ by regressing $\widehat{U}$ on $\widehat{V}$. 

\section{Generalization to other estimands} \label{sec:WATE}

This section generalizes the proposed method to estimands beyond ATT. Following \cite{li2018balancing}, we formulate estimands on various target populations via weighting. Specifically, let $f(x)$ be the covariates distribution in the observed data, which include both the external and concurrent data, and let $f(x)h(x)$ denote the covariate distribution of the target population, where $h(x)$---usually pre-specified---is a function that re-weights the concurrent population to the target population. We can represent the average treatment effect on the target population $f(x)h(x)$ as a weighted average treatment effect (WATE)   
\begin{equation}
    \tau^h = \frac{\mathbb{E}[h(X_i) (Y^{1}_i(1) - Y^{1}_i(0))]}{\mathbb{E}[h(X_i)]}. \label{eq:weightedATE}
\end{equation}
Several common estimands are special cases of the WATE. When $h(X) = e_Z(X)$, $\tau^h$ is the $\tau^{\ATT}$ as defined before. When $h(X) = 1-e_Z(X)$, $\tau^h=\mathbb{E}[Y^{1}_i(1) - Y^{1}_i(0)\mid Z_i=0]$, which is the average treatment effect on the population represented by the external study, akin to the average treatment effect for the control (ATC) in standard causal inference literature.  When $h(X) =1$, $\tau^{h} = \mathbb{E}[Y^{1}_i(1) - Y^{1}_i(0)],$
which is the average treatment effect on the combined population represented by both the concurrent and external studies, akin to the average treatment effect (ATE) estimand in standard causal inference literature. When $h(X) = e_Z(X)(1-e_Z(X))$,  $\tau^h$ is the average treatment effect on the overlapped population (ATO) between the concurrent and external studies. 

In all the above estimands, $h(x)$ is a function of $e_Z(x)$, which tilts the target population between the external and the concurrent study. With a slight abuse of notation, let the tilting function be $h(e_Z(x))$. Following the same analytical procedure in Section \ref{sec:unknown-b}, we derive a consistent and locally efficient estimator of $\tau^h$ as (the derivation is relegated to the Supplementary Material): 
\begin{equation} \label{eq:generalized-estimator}
    \widehat{\tau}^h = \frac{\sum_{i=1}^n  {h}(\widehat{e}_{Zi})[\widehat{\lambda}_i (\widehat{\mu}_{11i} - \widehat{\mu}_{10i}) + \widehat{T}_{i}]}{\sum_{i=1}^n {h}(\widehat{e}_{Zi}) \widehat{\lambda}_i},
\end{equation}
where 
\begin{equation}
    \widehat{T}_i = \frac{\widehat{R}_{11i}}{\widehat{e}_{Zi}\widehat{e}_{Ai}} - \frac{\widehat{R}_{10i} + \widehat{R}_{00i}}{1 - \widehat{e}_{Zi}\widehat{e}_{Ai}}, \quad \widehat{\lambda}_i = (Z_i - e_{Zi})\frac{h'(e_{Zi})}{h(e_{Zi})} + 1.
\end{equation}
The previous section focuses on the ATT, with $h(e_Z) = e_Z$, where \eqref{eq:generalized-estimator} reduces to $\widehat{\tau}^{\ATT}_{aug}$ in \eqref{eq:ATT-estimator-aug}. For the ATO estimand, estimator \eqref{eq:generalized-estimator} reduces to:
\begin{equation}
    \widehat{\tau}^{\ATO} = \frac{\sum_{i=1}^n  [Z_i(1 - 2\widehat{e}_{Zi}) + \widehat{e}_{Zi}^2](\widehat{\mu}_{11i} - \widehat{\mu}_{10i}) + \widehat{e}_{Zi} (1 - \widehat{e}_{Zi})\widehat{T}_{i}}{\sum_{i=1}^n Z_i(1 - 2\widehat{e}_{Zi}) + \widehat{e}_{Zi}^2}.
\end{equation}
The estimators of the ATE and ATC estimands are relegated to the supplementary material.

\section{Simulations} \label{sec:simulation}

We conduct simulations to examine the finite sample performance of the proposed estimators compared to existing estimators under various  settings. 

\subsection{Simulation 1: Homogeneous Treatment Effect with Constant $b(X) = b$} \label{sec:sim_1}

We simulate $N = 1000$ units independently. Each unit has four covariates $X = (X_1, X_2, X_3, X_4)'$: one binary with $X_1 \sim 2\mathsf{Ber}(0.5) - 1$ and three continuous with $X_2, X_3, X_4 \stackrel{iid}{\sim} \mathsf{N}(0, 1)$. All four covariates have mean 0 and standard deviation 1. We generate $Z \sim \mathsf{Ber}(e_Z)$ where $e_Z = \mathrm{expit}(\alpha_Z + X'\beta_Z)$. We set $\beta_Z = (-0.35, 0.3, 1.2, 0.5)'$ and adjust $\alpha_Z$ so that the number of units in the external data is equal to that in the concurrent data. Within $Z = 1$, we generate $A \sim \mathsf{Ber}(1 / (1 + m))$ so that within the concurrent data, the control arm and the treatment arm follow a $1:m$ allocation, with $m$ taking values $1, 2, 5, 10,$ and $20$. In other words, the proportion of units in the $(0,0), (1,0)$ and $(1,1)$ groups are $\pi = \left(0.5, 0.5/(1+m), 0.5m/(1+m)\right)$, respectively. We generate the potential outcomes from $Y^z(a) \sim \mathsf{N}(0.3 + bz + 0.4za + X'\beta_Y, 1)$, where $\beta_Y = (-0.4, 0.3, -0.7, -0.4)'$, and the observed outcomes $Y = (1 - Z)Y^0(0) + Z(1 - A)Y^1(0) + ZAY^1(1)$. Such a data generating process implies ME when $b = 0$ and homogeneous treatment effect with $Y_{i}^{1}(1) - Y_{i}^{1}(0) = 0.4$ for every unit. We range $b$ within $\{0, 0.2, 0.4\}$. For each setting, we repeat the process for $B = 1,000$ times and compare the following estimators with respect to the empirical bias and variance.

We focus on the target estimand $\tau^{\ATT}$, whose true value is 0.4. We obtain three augmented estimators, as defined in Equation \eqref{eq:ATT-estimator-aug}, with $b(X)$ being zero, constant, and flexible with no particular parametric form, denoted by $\widehat{\tau}_{aug}^{\mathrm{ME}}$, $\widehat{\tau}_{aug}^{\mathrm{const}}$ and $\widehat{\tau}_{aug}^{\mathrm{flex}}$, respectively.  
When $b(X) = 0$, $\widehat{\tau}_{aug}^{\mathrm{ME}}$ corresponds to the locally efficient estimator $\widehat{\tau}^{\mathrm{ATT}}_{aug}$ proposed by \cite{li2023improving, gao2023integrating} that assumes ME, under which $\widehat{\mu}_{10} = \widehat{\mu}_{00}$ are estimated by a single linear regression with the main effects of the four covariates in the combined $(1, 0)$ and $(0, 0)$ groups. When $b(X)$ is assumed to be constant, $Z$, is added to the linear regression. Finally, when we do not impose any parametric specification on $b(X)$, $\widehat{\mu}_{10}$ and $\widehat{\mu}_{00}$ are estimated separately by linear regression with the main effects of the four covariates in the $(1, 0)$ and $(0, 0)$ groups, respectively. The three estimators $\widehat{\tau}_{aug}^{\mathrm{ME}}$, $\widehat{\tau}_{aug}^{\mathrm{const}}$ and $\widehat{\tau}_{aug}^{\mathrm{flex}}$ offer increasing flexibility in the specification of $b(X)$.

We compare the bias and standard deviation of the above three estimators with alternatives that are commonly used in practice for related applications. First, as a benchmark for the randomized comparison without external data we consider (a) the mean difference (MD) estimator within the concurrent RCT, 
\begin{equation}
    \widehat{\tau}_{\mathrm{MD}} = \frac{\sum_{i=1}^n Z_i A_i Y_i}{\sum_{i=1}^n Z_iA_i} - \frac{\sum_{i=1}^n Z_i (1 - A_i) Y_i}{\sum_{i=1}^n Z_i (1 - A_i)};
\end{equation}
and (b) the mean difference in model-based predicted outcomes (MDP) within the RCT  $\widehat{\tau}_{\mathrm{MDP}} = \sum_{i=1}^N Z_i(\widehat{\mu}_{11i} - \widehat{\mu}_{10i}) / N_1$ with $\widehat{\mu}_{11i}$ and $\widehat{\mu}_{10i}$ estimated from a correctly specified regression model, fit to the RCT.

Alternative estimators that incorporate the EC data are:
(a) a propensity score (PS) weighted estimator,
\begin{equation}
    \widehat{\tau}_{\mathrm{PS}} = \frac{\sum_{i=1}^n Z_i A_i Y_i \widehat{w}_i}{\sum_{i=1}^n Z_iA_i \widehat{w}_i} - \frac{\sum_{i=1}^n (1 - Z_iA_i) Y_i\widehat{w}_i}{\sum_{i=1}^n (1 - Z_iA_i)\widehat{w}_i},
\end{equation}
where $\widehat{w}_i = Z_i + (1 - Z_i) \frac{\widehat{e}_{Zi}}{1 - \widehat{e}_{Zi}}$ reweights the external control to resemble the RCT; (b) the doubly robust (DR) estimator 
\begin{equation}\label{eq:ATT-estimator-dr}
    \widehat{\tau}_{\mathrm{DR}} = \frac{1}{N_1} \sum_{i=1}^N Z_i\left\{ \widehat{\mu}_{11i} - \widehat{\mu}_{10i} + \frac{1}{\widehat{e}_{Ai}} A_i(Y_i - \widehat{\mu}_{11i}) - \dfrac{1}{1 - \widehat{e}_{Ai}}(1 - A_i)(Y_i - \widehat{\mu}_{10i})\right\},
\end{equation} where $\widehat{\mu}_{11i}$ and $\widehat{\mu}_{10i}$ are the same as in $\widehat{\tau}_{aug}^\mathrm{ME}$; (c) the ANCOVA estimator assuming ME, $\widehat{\tau}_{\mathrm{ANCOVA}}^\mathrm{ME}$, the estimated coefficient of $A$ in linear regression $Y \sim X + A$ using all units; (d) another ANCOVA estimator allowing for constant $b(X)$, $\widehat{\tau}_{\mathrm{ANCOVA}}^{\mathrm{const}}$, the estimated coefficient of $A$ in linear regression $Y \sim X + A + Z$ using all units.   In all these approaches, the propensity scores are estimated by logistic regressions with main effects of the four covariates.

In Table \ref{tab:sim_1} the difference-in-mean estimator $\widehat{\tau}_{\mathrm{MD}}$ is unbiased, as expected, regardless of the data generating process. It also has larger standard deviation than all other estimators, since it does not utilize any information from the covariates and only uses data in the concurrent RCT. $\widehat{\tau}_{\mathrm{MDP}}$ is also unbiased and slightly more efficient, showing the best that one could obtain from modeling the outcome in the RCT alone. 
These serve as a benchmark to compare external augmentation. The propensity score weighted estimator $\widehat{\tau}_{\mathrm{PS}}$ incorporates data from the external control via weighting, and thus has smaller standard error than $\widehat{\tau}_{\mathrm{MD}}$; however, $\widehat{\tau}_{\mathrm{PS}}$ is biased when $b \neq 0$. The remaining estimators require specification of the outcome models. The augmented estimator $\widehat{\tau}_{aug}^{\mathrm{flex}}$ allows for the most flexibility and does not improve upon the model-based RCT benchmark $\widehat{\tau}_{\mathrm{MD}}$.  The augmented estimator $\widehat{\tau}_{aug}^{\mathrm{const}}$ assumes that $b(X)$ is a constant, which aligns with the true data generating process. Therefore, it is unbiased and has slightly smaller standard deviation than $\widehat{\tau}_{aug}^{\mathrm{flex}}$. The ANCOVA estimator $\widehat{\tau}_{\mathrm{ANCOVA}}^{\mathrm{const}}$ involves analogous assumptions and performs similarly to $\widehat{\tau}_{aug}^{\mathrm{const}}$. The remaining estimators,  $\widehat{\tau}_{aug}^{\mathrm{ME}}$ and  $\widehat{\tau}_{\mathrm{ANCOVA}}^{\mathrm{ME}}$, both assume ME leading to much smaller standard errors than the other estimators, but they are biased when $b \neq 0$. The doubly robust estimator $\widehat{\tau}_{\mathrm{DR}}$, in contrast, remains unbiased even though the outcome model relies on the incorrect ME assumption. This is due to the double robustness with the correctly specified propensity score model at the expense of increased standard error, which is no better than the model-based RCT benchmark $\widehat{\tau}_{\mathrm{MD}}$.

Table \ref{tab:sim_1} shows that the efficiency gain of some estimators over $\widehat{\tau}_{\mathrm{MD}}$ is highly dependent on additional assumptions on $b(X)$, especially ME, explicitly or implicitly. These estimators should be used with strong scrutiny of the ME assumption. When the ME assumption is violated, they can lead to largely biased estimates with a small standard deviation and result in over-confident conclusions.

\begin{table}[h]
    \centering
    \resizebox{\textwidth}{!}{
        \begin{tabular}{cccccccccccccc}
            \toprule
            & & \multicolumn{2}{c}{RCT only} & \multicolumn{7}{c}{With external control}\\
            \cmidrule(lr){3-4} \cmidrule(lr){5-11}
            $b$ & 1:$m$ & $\widehat{\tau}_{\mathrm{MD}}$ & $\widehat{\tau}_{\mathrm{MDP}}$ & $\widehat{\tau}_{\mathrm{PS}}$ & $\widehat{\tau}_{\mathrm{DR}}$ & $\widehat{\tau}_{\mathrm{ANCOVA}}^{\mathrm{ME}}$ & $\widehat{\tau}_{\mathrm{ANCOVA}}^{\mathrm{const}}$ & $\widehat{\tau}_{aug}^{\mathrm{ME}}$ & $\widehat{\tau}_{aug}^{\mathrm{const}}$ & $\widehat{\tau}_{aug}^{\mathrm{flex}}$\\ 
            \midrule
            \multirow{5}{*}{0} & 1:1 & -1 (10) & 0 (6) & -1 (11) & 0 (6) & 0 (6) & 0 (6) & 0 (6) & 0 (6) & 0 (6) \\
             & 1:2 & 0 (11) & 0 (7) & -1 (11) & 0 (7) & 0 (6) & 0 (7) & 0 (6) & 0 (7) & 0 (7) \\
             & 1:5 & 0 (15) & 0 (10) & -1 (11) & 0 (10) & 0 (6) & 0 (10) & 0 (7) & 0 (10) & 0 (10) \\
             & 1:10 & 0 (21) & 0 (16) & -1 (12) & 0 (16) & 0 (7) & 0 (15) & 0 (8) & 0 (15) & 0 (16) \\
             & 1:20 & -1 (29) & -1 (24) & -1 (12) & -1 (23) & 0 (7) & -1 (21) & 0 (9) & -1 (22) & -1 (24) \\
            \midrule
            \multirow{5}{*}{20} & 1:1 & -1 (10) & 0 (6) & 13 (11) & 0 (6) & 11 (6) & 0 (6) & 10 (6) & 0 (6) & 0 (6) \\
             & 1:2 & 0 (11) & 0 (7) & 14 (11) & 0 (7) & 13 (6) & 0 (7) & 11 (6) & 0 (7) & 0 (7) \\
             & 1:5 & 0 (15) & 0 (10) & 16 (11) & 0 (10) & 16 (6) & 0 (10) & 14 (7) & 0 (10) & 0 (10) \\
             & 1:10 & 0 (21) & 0 (16) & 17 (12) & 0 (16) & 17 (7) & 0 (15) & 16 (8) & 0 (15) & 0 (16) \\
             & 1:20 & -1 (29) & -1 (24) & 18 (12) & -1 (23) & 18 (7) & -1 (21) & 17 (9) & -1 (22) & -1 (24) \\
            \midrule
            \multirow{5}{*}{40} & 1:1 & -1 (10) & 0 (6) & 26 (11) & 0 (6) & 22 (6) & 0 (6) & 20 (6) & 0 (6) & 0 (6) \\
             & 1:2 & 0 (11) & 0 (7) & 29 (11) & 0 (7) & 26 (6) & 0 (7) & 23 (6) & 0 (7) & 0 (7) \\
             & 1:5 & 0 (15) & 0 (10) & 34 (12) & 0 (10) & 31 (7) & 0 (10) & 28 (7) & 0 (10) & 0 (10) \\
             & 1:10 & 0 (21) & 0 (16) & 36 (12) & 0 (16) & 35 (7) & 0 (15) & 32 (8) & 0 (15) & 0 (16) \\
             & 1:20 & -1 (29) & -1 (24) & 37 (12) & -1 (24) & 37 (7) & -1 (21) & 34 (9) & -1 (22) & -1 (24) \\
            \bottomrule\\
        \end{tabular}
    }

    \caption{Bias and standard deviation (in brackets) of the estimators described in Section \ref{sec:sim_1} under different values of the systematic difference $b$ and treatment-to-control ratio $m$ in Simulation 1 under homogeneous treatment effects. The estimators include two estimators that only use the RCT data: the mean difference $\widehat{\tau}_{\mathrm{MD}}$ and the mean difference in model-based predicted outcome $\widehat{\tau}_{\mathrm{MDP}}$, and seven estimators that utilize external control, including the propensity score weighted estimator $\widehat{\tau}_{\mathrm{PS}}$, the doubly robust estimator $\widehat{\tau}_{\mathrm{DR}}$, the ANCOVA estimators $\widehat{\tau}_{\mathrm{ANCOVA}}^{\mathrm{ME}}$ and $\widehat{\tau}_{\mathrm{ANCOVA}}^{\mathrm{const}}$ without and with a separate intercept for $Z$ respectively,
    and the proposed augmented estimators, $\widehat{\tau}_{\mathrm{aug}}^{\mathrm{ME}}$, $\widehat{\tau}_{\mathrm{aug}}^{\mathrm{const}}$ and $\widehat{\tau}_{\mathrm{aug}}^{\mathrm{flex}}$ with increasing flexibility of $b(x)$. All columns except 1:$m$ are multiplied by 100 and rounded to the nearest integer.}
    \label{tab:sim_1}
\end{table}

\subsection{Simulation 2: Heterogeneous Treatment Effect with $b(X)$ Dependent on $X$}\label{sec:sim_2}

We repeat Simulation 1 with different outcome models for $Y^z(a) \sim \mathsf{N}(\alpha_{za} + X'\beta_{za}, 1)$, where $(\alpha_{00}, \alpha_{10}, \alpha_{11}) = (0.3-b, 0.3, 0.7)$, $\beta_{00} = (-0.4-b, 0.4+2b, -0.7-b, -0.4-1.5b)'$, $\beta_{10} = (-0.4, 0.4, -0.7, -0.4)'$ and $\beta_{11} = (-0.8, 0.1, -0.5, -1.1)$. Here, $b(X) = b(1 + X_1 - 2X_2 + X_3 + 1.5X_4)$. 

We report the empirical estimate of the bias and variance of each of the estimators in Table \ref{tab:sim_2}. The pattern is similar to that in Simulation 1. The difference-in-mean estimator $\widehat{\tau}_{\mathrm{MD}}$ is unbiased with largest standard deviation and the propensity score weighted estimator $\widehat{\tau}_{\mathrm{PS}}$ is generally biased when $b \neq 0$. The estimators $\widehat{\tau}_{aug}^{\mathrm{flex}}$ and $\widehat{\tau}_{\mathrm{MDP}}$ do not assume any additional functional form of $b(X)$ and are unbiased. The estimators $\widehat{\tau}_{aug}^{\mathrm{const}}$ and $\widehat{\tau}_{\mathrm{ANCOVA}}^{\mathrm{const}}$ assume $b(X)$ to be a constant, which is not true under this setting, but their bias is still close to 0. In fact, $\widehat{\tau}_{\mathrm{ANCOVA}}^{\mathrm{const}}$ is guaranteed to be unbiased (see Supplementary Material for the proof). Whether $\widehat{\tau}_{aug}^{\mathrm{const}}$ will be more efficient than $\widehat{\tau}_{aug}^{\mathrm{flex}}$ reflects a trade-off between fewer model parameters and better predictability of the models. We replicate the same simulation but reduce the standard deviation of the error term in $Y$ from 1 to 0.2, where the covariates are much more predictive of the outcome (the results are relegated to the Supplementary Material). In that case, $\widehat{\tau}_{aug}^{\mathrm{const}}$  has larger standard deviation than $\widehat{\tau}_{aug}^{\mathrm{flex}}$, which uses correctly specified outcome models that utilize the high predictability. In contrast, $\widehat{\tau}_{aug}^{\mathrm{ME}}$ and $\widehat{\tau}_{\mathrm{ANCOVA}}^{\mathrm{ME}}$ are biased when ME does not hold, although they lead to lower standard deviation compared to other estimators.

\begin{table}[h]
    \centering
    \resizebox{\textwidth}{!}{
        \begin{tabular}{cccccccccccccc}
            \toprule
            & & \multicolumn{2}{c}{RCT only} & \multicolumn{7}{c}{With external control}\\
            \cmidrule(lr){3-4} \cmidrule(lr){5-11}
            $b$ & 1:$m$ & $\widehat{\tau}_{\mathrm{MD}}$ & $\widehat{\tau}_{\mathrm{MDP}}$ & $\widehat{\tau}_{\mathrm{PS}}$ & $\widehat{\tau}_{\mathrm{DR}}$ & $\widehat{\tau}_{\mathrm{ANCOVA}}^{\mathrm{ME}}$ & $\widehat{\tau}_{\mathrm{ANCOVA}}^{\mathrm{const}}$ & $\widehat{\tau}_{aug}^{\mathrm{ME}}$ & $\widehat{\tau}_{aug}^{\mathrm{const}}$ & $\widehat{\tau}_{aug}^{\mathrm{flex}}$\\ 
            \midrule
            \multirow{5}{*}{0} & 1:1 & -1 (12) & 0 (6) & -1 (12) & 0 (6) & 0 (7) & 0 (7) & 0 (6) & 0 (6) & 0 (6) \\
             & 1:2 & 0 (13) & 0 (7) & -1 (12) & 0 (7) & 1 (7) & 0 (7) & 0 (6) & 0 (7) & 0 (7) \\
             & 1:5 & 0 (16) & 0 (10) & -1 (12) & 0 (10) & 1 (7) & 0 (11) & 0 (7) & 0 (10) & 0 (10) \\
             & 1:10 & -1 (22) & 0 (16) & -1 (12) & 0 (16) & 2 (7) & 0 (16) & 0 (8) & 0 (15) & 0 (16) \\
             & 1:20 & -1 (30) & -1 (24) & -1 (12) & -1 (23) & 2 (8) & -1 (23) & 0 (9) & -1 (22) & -1 (24) \\
            \midrule
            \multirow{5}{*}{20} & 1:1 & -1 (12) & 0 (6) & 17 (13) & 0 (6) & 11 (7) & 0 (7) & 11 (6) & -1 (7) & 0 (6) \\
             & 1:2 & 0 (13) & 0 (7) & 20 (13) & 0 (7) & 14 (7) & 0 (8) & 13 (7) & -1 (8) & 0 (7) \\
             & 1:5 & 0 (16) & 0 (10) & 23 (14) & 0 (10) & 17 (7) & 0 (11) & 17 (7) & -1 (11) & 0 (10) \\
             & 1:10 & -1 (22) & 0 (16) & 24 (14) & 0 (16) & 19 (8) & 0 (17) & 20 (8) & -1 (17) & 0 (16) \\
             & 1:20 & -1 (30) & -1 (24) & 25 (15) & -1 (24) & 20 (8) & -1 (24) & 22 (9) & -2 (24) & -1 (24) \\
            \midrule
            \multirow{5}{*}{40} & 1:1 & -1 (12) & 0 (6) & 35 (15) & 0 (6) & 22 (8) & 0 (8) & 21 (7) & -1 (7) & 0 (6) \\
             & 1:2 & 0 (13) & 0 (7) & 40 (16) & 0 (7) & 26 (8) & 0 (9) & 26 (7) & -1 (9) & 0 (7) \\
             & 1:5 & 0 (16) & 0 (10) & 46 (16) & 0 (11) & 33 (8) & 0 (13) & 34 (8) & -2 (13) & 0 (10) \\
             & 1:10 & -1 (22) & 0 (16) & 49 (17) & 0 (16) & 36 (8) & 0 (19) & 40 (9) & -2 (21) & 0 (16) \\
             & 1:20 & -1 (30) & -1 (24) & 51 (18) & 0 (25) & 39 (9) & -1 (26) & 44 (11) & -2 (30) & -1 (24) \\
            \bottomrule\\
        \end{tabular}
    }
    \caption{Bias and standard deviation (in brackets) of the estimators described in Section \ref{sec:sim_2} under different values of the systematic difference $b$ and treatment-to-control ratio $m$ in Simulation 2 under heterogeneous treatment effects. The estimators include two estimators that only use the RCT data: the mean difference $\widehat{\tau}_{\mathrm{MD}}$ and the mean difference in model-based predicted outcome $\widehat{\tau}_{\mathrm{MDP}}$, and seven estimators that utilize external control, including the propensity score weighted estimator $\widehat{\tau}_{\mathrm{PS}}$, the doubly robust estimator $\widehat{\tau}_{\mathrm{DR}}$, the ANCOVA estimators $\widehat{\tau}_{\mathrm{ANCOVA}}^{\mathrm{ME}}$ and $\widehat{\tau}_{\mathrm{ANCOVA}}^{\mathrm{const}}$ without and with a separate intercept for $Z$ respectively,
    and the proposed augmented estimators, $\widehat{\tau}_{\mathrm{aug}}^{\mathrm{ME}}$, $\widehat{\tau}_{\mathrm{aug}}^{\mathrm{const}}$ and $\widehat{\tau}_{\mathrm{aug}}^{\mathrm{flex}}$ with increasing flexibility of $b(x)$. All columns except 1:$m$ are multiplied by 100 and rounded to the nearest integer.}
    \label{tab:sim_2}
\end{table}

\subsection{Simulation 3: Non-linear outcome model}\label{sec:sim_3}

In this simulation, we demonstrate the common case where the outcomes $\mathbb{E}[Y^z(a)\mid X]$ is not linear in $X$ but are fit with linear outcome models instead. We repeat Simulation 2, but differently, we generate the potential outcomes from $Y^0(0) \sim \mathsf{N}(0.3 - b - (0.4+b)X_1 + (0.4+2b)X_2 - (0.7+b)X_3 - (0.4+1.5b)X_4 + (0.9+b)(X_2^2 - 1), 1)$, $Y^1(0) \sim \mathsf{N}(0.3 - 0.4X_1 + 0.4X_2 - 0.7X_3 - 0.4X_4 + 0.9(X_2^2 - 1), 1)$ and $Y^1(1) \sim \mathsf{N}(0.7 - 0.8X_1 + 0.1X_2 - 0.5X_3 - 1.1X_4 + 0.6(X_2^2 - 1), 1)$. Here, $b(X) = b(1 + X_1 - 2X_2 + X_3 + 1.5X_4 - (X_2^2 - 1))$.

We observe similar patterns to Simulation 2, and hence we relegate the empirical estimate of the bias and variance of each of the five estimators to the Supplementary Material. 

\subsection{Simulation 4: Other Estimands}\label{sec:sim_4}

We simulate data as in Simulation 2 with heterogeneous treatment effects. We use \eqref{eq:ATT-estimator-aug} to estimate the ATE, ATC and ATO estimands. For each estimand, we estimate the outcome by linear regression on the main effect of the four covariates, with three different assumptions on $b(X)$: (1) mean exchangeability $b(X) = 0$; (2) $b(X)$ is constant; (3) no particular form of $b(X)$. The variance and bias of each estimator is shown in the supplemetary material. 

The estimator $\widehat{\tau}^{\mathrm{ME}}_{*}$ is generally biased when ME does not hold, for $* \in \{\mathrm{ATE}, \mathrm{ATC}, \mathrm{ATO}\}$. Different from the ATT estimand, the constant $b(X)$ assumption no longer always leads to unbiased estimators. The estimator $\widehat{\tau}^{\mathrm{flex}}_{*}$ is still unbiased, but with larger standard deviation. 

We conduct a similar simulation where $b(X)$ is a constant, with the results relegated to the supplement material. Under constant $b(X)$, $\widehat{\tau}^{\mathrm{const}}_{*}$ is unbiased and the reduction in standard deviation from $\widehat{\tau}^{\mathrm{flex}}_{*}$ is more pronounced than ATT.

\section{Application} \label{sec:application}
We illustrate the proposed method using two clinical trials of type 2 diabetes patients and cardiovascular outcomes: TECOS \citep{green2015effect} and EXSCEL \citep{holman2017effects}. The active treatment is sitagliptin in TECOS and exenatide in EXSCEL. These trials had similar populations, identical endpoints, and a sizable placebo arm. This allows us to test methods for trial augmentation in a setting where a gold-standard, internal placebo arm is available. The existence of internal control data also allows the proposed methods to be applied, with and without the assumption of ME.  

The primary endpoint is a composite of adjudicated cardiovascular death, myocardial infarction and stroke. We analyze the 1-year cardiovascular composite endpoint as a binary-outcome. Although most inclusion and exclusion criteria were similar, EXSCEL was more inclusive, and we applied common criteria prior to conducting analysis (Supplementary Material).  This reduced the original samples sizes from 14,752 and 14,671 to 3,836 and 9,475, in EXSCEL and TECOS respectively. The substantial drop in EXSCEL is attributable to requiring prior cardiovascular disease and Hba1c value less than 8, as in TECOS. We further limited the data sources to countries with representation in both trials, to avoid perfect confounding with country enrollment. Factors related to cardiovascular risk were collected similarly in both studies and previously harmonized to have nearly identical definitions. Variables used in this analysis were pre-specified including demographics, vital signs, lab values, prior cardiovascular events, comorbidities, and cardiovascular medications (Supplementary Material). These were used for adjustment in outcome and propensity score models. 

Using the proposed methods we conduct two separate analyses, one in each RCT augmented by the external control data from the other trial. Specifically, we conduct the analysis of TECOS, comparing 4381 patients randomized to sitagliptan to 4404 randomized to placebo control, augmented by an additional 1795 placebo control patients from EXSCEL. We then conduct the analysis of EXSCEL, comparing 1748 patients randomized to exenatide to 1795 randomized to placebo control, augmented by an additional 4404 placebo control patients from TECOS. Results are shown in Table \ref{combined_results}.

\begin{table}[h]
\centering
\begin{tabular}{|l|c|c|c|l|c|c|c|}
\toprule
&\multicolumn{3}{|c||}{TECOS trial + EXSCEL Placebo}& & \multicolumn{3}{c|}{EXSCEL trial + TECOS Placebo} \\
\hline
Method & Estimate & Lower CI & Upper CI &  & Estimate & Lower CI & Upper CI \\
\hline
$\widehat{\tau}_{\mathrm{MD}}$ & -0.8 & -1.9 & 0.4 & & -1.2 & -3.2 & 0.8 \\
$\widehat{\tau}_{\mathrm{aug}}^{\mathrm{ME}}$ & -1.2 & -2.3 & -0.1 &   & 0.5 & -0.9 & 2.0 \\
$\widehat{\tau}_{\mathrm{aug}}^{\mathrm{const}}$ & -0.6 & -1.7 & 0.5 &  & -0.9 & -2.7 & 0.9 \\
$\widehat{\tau}_{\mathrm{aug}}^{\mathrm{flex}}$ & -0.6 & -1.7 & 0.5  &    & -0.9 & -2.7 & 1.0 
\\
        \bottomrule \\
    \end{tabular}
\caption{Analysis Results for TECOS and EXSCEL trials. Estimate: difference in the percentage of composite endpoints with Lower and Upper Confidence intervals (CI). Method: $\widehat{\tau}_{\mathrm{MD}}$ is the randomized comparison, after applying exclusions for data harmonization, and without any external augmentation ($N$=9,475 in TECOS; $N$=3,836 in EXSCEL); $\widehat{\tau}_{\mathrm{aug}}^{\mathrm{ME}}$ corresponds to the proposed method assuming ME where $b(X) = 0$; $\widehat{\tau}_{\mathrm{aug}}^{\mathrm{const}}$ corresponds to the proposed method assuming  $b(X;\theta) = \theta$; $\widehat{\tau}_{\mathrm{aug}}^{\mathrm{flex}}$ corresponds to the proposed method assuming  $b(X; \theta) = \theta_0 + X'\theta_1$.}
\label{combined_results}
\end{table}

First, considering the randomized comparison within TECOS, prior to augmentation, there is a non-significant reduction of 0.80 in the percentage of patients with a composite endpoint associated with treatment. The confidence interval includes a difference as large as -1.9 or 0.4 in the other direction. When the analysis is augmented under the assumption of ME the result changes. The point estimate is -1.2 and the confidence interval does not contain 0. This suggests a treatment benefit, but could also be explained by bias attributable to differences between EXSCEL and TECOS.  Allowing for those differences, with a constant intercept term or linear heterogeneity, the apparent benefit disappears (Table \ref{combined_results}). 

Next, considering the randomized comparison within EXSCEL, prior to augmentation, there is a non-significant reduction of 1.2 in the percentage of patients with a composite endpoint associated with treatment.  The augmented result under mean eachangeability switches direction, but remains non-significant. Relaxing that assumption, with either a constant or a linear model, allows the randomized treatment effect to be recovered quite closely  (Table \ref{combined_results}).  

\section{Discussion} \label{sec:discussion}

When RCTs are infeasible or limited by sample size, augmentation with EC data becomes attractive. In this paper we introduce a double index notation to account for the possibility that different versions of the outcome may arise in RCT vs EC data. Using the double index notation, we establish causal estimands and identifying assumptions. Among the assumptions, ME is unique to this context and violated if either unconfounded trial participation (Assumption \ref{ass:unconfoundZ}) or ME-PO (Assumption \ref{ass:mean-exchangeability}) does not hold.  We propose a locally efficient semiparametric estimator that combines concurrent and external data without assuming ME. Simulation and illustration in EXSCEL and TECOS shows that this estimator is more robust than traditional approaches. However, there is an tradeoff between efficiency and robustness; whereby the assumption of ME, if valid, leads to greater efficiency. These tradeoffs must be considered in trials augmented by ECs. The proposed method may be used for primary analysis or for sensitivity analysis in augmented trials. 

The double index notation proposed here focuses on the possibility that RCT and EC data have different versions of outcome. Another source of difference could arise from trial participation effects, or the Hawthorne effects, by which engagement in a study impacts outcome even when treatment is fixed (e.g. placebo). Alternative notation would incorporate trial participation effects into a dual potential outcome, as in $Y_i(z,a)$ \citep{li2023improving, ung2024generalizing}.  The index $z$ implies that there is a different potential outcome for the same individual, $i$, if that individual participates in the trial versus does not participate in the trial, for fixed treatment $a$. Hawthorne effects can often be excluded by substantive knowledge \citep{li2023improving}, such that $Y_i(z,a) = Y_i(a)$. If so, the extra notation is unnecessary and is generally not used.  Our notation could be expanded to permit both different versions of outcome and Hawthorne effects as $Y^{z}(z,a)$. This distinction does not change the proposed estimator but clarifies the source(s) of bias that arise if $b(X)$ is not estimated but assumed to be zero. 
 
RCTs augmented by EC data are often viewed as an observational treatment comparison and analyzed using propensity score methods. Indeed, there are analogous assumptions. Observational treatment comparisons make the assumption of no unmeasured confounding, wherein the measured patient characteristics capture all of the important differences between non-randomized treatment arms.  Assumption \ref{ass:unconfoundZ} is analogous for the RCT versus EC. This assumption will hold if the data contain sufficient covariates to measure the differences between patients who enroll in the RCT versus EC. While this assumption is analogous, the consequence of unmeasured confounding could be more serious for an RCT augmented by EC, where regulatory or payer decisions depend strongly on the result. In such cases it is uniquely important for the study design to include some randomized concurrent control data, rather than conduct single arm trials. Leveraging concurrent control data, the proposed method is robust to violations of Assumption \ref{ass:unconfoundZ}.  

Robustness to ME comes at a cost to precision.  In the EXSCEL and TECOS example, the robust augmented estimator was only slightly more precise than the randomized trial alone.  This is consistent with the simulation results where the augmented estimators provided substantial improvement in precision only when randomized trial had very few controls. The concurrent control arm in EXSCEL and TECOS is already so large that gains from augmentation are small. That is both an advantage, in terms of having a gold standard, but a disadvantage for the value of augmentation.  In addition, the added precision may depend on how well the measured covariates predict outcome, as the model for outcome is the source of shared information.  Future directions for research would evaluate how results depend on the selection of covariates for the outcome model and whether greater efficiency can be achieved through alternative modeling techniques. For example, machine learning or shrinkage methods could be used to select the most important variables from a large set of candidates. This selection step would generally be unacceptable if covariates were required to justify unconfounded trial participation, Assumption \ref{ass:unconfoundZ}. However the proposed estimator relaxes this assumption, and therefore may permit efficient use of covariates without increasing bias. 

In Section \ref{sec:WATE} we introduce weighted causal treatment effect estimands and extend the proposed estimator to these target effects. Elsewhere, we have addressed the conceptual model and interpretation of weighted average causal effects \citep{wang2025integratingrandomizedcontrolledtrial} for RCTs augmented by EC.  The extensive literature on generalizability supports the idea that EC populations are often a more appropriate target population than RCTs \citep{stuart2011use,dahabreh2019generalizing, dahabreh2021study, colnet2024causal, ung2024generalizing}.  If so, the causal treatment effect estimand would target the EC (ATC in Section \ref{sec:WATE}).  This differs from the literature on RCTs augmented by ECs, which often target the RCT population \citep{li2023improving, cheng2023enhancing, van2024adaptive}.  With the recognition that neither population is correct, and both are relevant to generalizability, one might prefer to develop a principled mixture\citep{wang2025integratingrandomizedcontrolledtrial}. Section \ref{sec:WATE} provides a number of relevant weighted estimands and estimators that integrate the RCT and EC target population. 
 
\section*{Acknowledgments}
Laine E Thomas, Ph.D., holds an Innovation in Regulatory Science Award from the Burroughs Wellcome Fund. The contents of this article are solely the responsibility of the authors and do not necessarily represent the view of BWF. We thank the EXSCEL and TECOS publications committees, Michael Pencina, Hwanhee Hong, Yanxun Xu for insightful discussions, and Dylan Thibault for data analysis. The TECOS trial was funded by Merck Sharp \& Dohme Corp., a subsidiary of Merck \& Co., Inc. (Kenilworth, NJ). The EXSCEL trial was sponsored and funded by Amylin Pharmaceuticals Inc. (San Diego, CA), a wholly owned subsidiary of AstraZeneca (Gaithersburg, MD).   

\bibliographystyle{biom}
\bibliography{overlap.bib}
 
\section*{Supplementary Materials}

Web Appendices, Tables, and Figures referenced in Sections \ref{sec:no-exchange} to \ref{sec:application} are available with this paper at the Biometrics website on Oxford Academic.

\end{document}


\date{{\it Received XXX} 20XX. {\it Revised XXX} 20XX.  {\it
Accepted XXX} 20XX.}



\pagerange{\pageref{firstpage}--\pageref{lastpage}} 
\volume{XX}
\pubyear{20XX}
\artmonth{XXX}


\doi{XXX}

\label{firstpage}

\maketitle
\section{Proof of Equation (9) in Section 4}

\begin{align*}
    \bE[Y^1(0)\mid X]\stackrel{Ass~2}{=}&\bE[Y^1(0)\mid X, Z=1]\\
                \stackrel{Ass~3}{=}&\bE[Y^1(0)\mid X, Z=1, A=0]\\
                \stackrel{Ass~1}{=}&\bE[Y\mid X, Z=1, A=0].
\end{align*}
\begin{align*}
        \bE[Y^0(0)\mid X]\stackrel{Ass~2}{=}&\bE[Y^0(0)\mid X, Z=0]\\
                \stackrel{design}{=}&\bE[Y^0(0)\mid X, Z=0, A=0]\\
                \stackrel{Ass~1}{=}&\bE[Y\mid X, Z=0, A=0].
\end{align*}

\section{Moment of estimating equation (10) in Section 4}

We show that the estimating equation satisfies the moment condition:
\begin{equation}
    \mathbb{E}\left[C(X)\left\{Z(1-A)[Y - \mu_{10}(X; \beta) + (1 - Z)[Y + b(X) - \mu_{10}(X;\beta)\right\}\right] = 0.
\end{equation}
To see this, note that 
\begin{align*}
    \mathbb{E}[Z(1-A)[Y - \mu_{10}(X;\beta)] \mid X] &= \mathbb{E}[Z(1-A)\mathbb{E}[Y - \mu_{10}(X;\beta) \mid X, Z, A] \mid X] \\
    &= e_Z(X)(1 - e_A(X))\left\{\mathbb{E}[Y \mid X, Z = 1, A = 0] - \mu_{10}(X; \beta)\right\} \\
    &=e_Z(X)(1 - e_A(X))\left\{\mu_{10}(X; \beta) - \mu_{10}(X; \beta)\right\} \\
    &= 0,
\end{align*}
and 
\begin{align*}
    \mathbb{E}[(1-Z)[Y + b(X) - \mu_{10}(X;\beta)] \mid X] &= \mathbb{E}[(1 - Z)\mathbb{E}[Y + b(X) - \mu_{10}(X;\beta) \mid X, Z] \mid X] \\
    &= (1 - e_Z(X))\left\{\mathbb{E}[Y \mid X, Z = 0] + b(X) - \mu_{10}(X; \beta)\right\} \\
    &=(1 - e_Z(X))\left\{\mu_{00}(X; \beta) + b(X) - \mu_{10}(X; \beta)\right\} \\
    &= 0.
\end{align*}
Therefore,
\begin{align*}
    \text{LHS} &= \mathbb{E}\left[C(X)\{\mathbb{E}[Z(1-A)[Y-\mu_{10}(X;\beta)] \mid X] + \mathbb{E}[(1-Z)[Y + b(X) - \mu_{10}(X;\beta)\mid X]\}\right] \\
    &= 0.
\end{align*}

\section{Derivation of $\widehat{\tau}^h$ in Section 5}

We first show that the EIF for $\tau^h$ is given by 
\begin{align}
    \mathrm{IF}(Y, X, Z, A) = &\frac{h(e_Z(X))}{\mathbb{E}[h(e_Z(X)]}\Bigg\{
    \lambda(\Delta(X) - \tau^h) + \frac{ZA}{e_Z(X) e_A(X)}[Y - \mu_{11}(X)] \nonumber \\
    &- \frac{Z(1 - A)[Y - \mu_{10}(X)] + r(X)(1-Z)[Y - \mu_{00}(X)]}{e_Z(X)[1 - e_A(X)] + (1 - e_Z(X)) r(X)}
    \Bigg\}, \label{eq:EIF}
\end{align}
where 
\begin{equation}
    \lambda = (Z - e_Z(X))\frac{\dot{h}(e_Z(X))}{h(e_Z(X))} + 1, \quad r(X) = \frac{\mathrm{Var}[Y^1(0) \mid X]}{\mathrm{Var}[Y^0(0) \mid X]}.
\end{equation}

Denote the density of $X$ by $p(x)$ and denote the conditional density $p(Y^z(a) = y\mid Z = z, A = a)$ by $p_{za}(y \mid x)$. The full observational data distribution is then 
\begin{equation}
    p(y, x, z, a) = f(x)(1 - e_Z(x))^{1 - z}e_Z(x)^z(1-e_A(x))^{(1-a)z}e_A(x)^{az}p_{00}(y \mid x)^{1-z} p_{10}(y \mid x)^{(1-a)z} p_{11}(y \mid x)^{az}.
\end{equation}
We consider a regular parametric submodel indexed by $\varphi$ with true value $\varphi^*$:
\begin{align}
    p(y, x, z, a; \varphi) =& f(x; \varphi)(1 - e_Z(x; \varphi))^{1 - z}e_Z(x; \varphi)^z(1-e_A(x; \varphi))^{(1-a)z}e_A(x; \varphi)^{az} \nonumber \\  &\cdot
    p_{00}(y \mid x; \varphi)^{1-z} p_{10}(y \mid x; \varphi)^{(1-a)z} p_{11}(y \mid x; \varphi)^{az},
\end{align}
where 
\begin{align}
    b(x; \varphi) &= \mathbb{E}[Y^1(0) \mid Z = 1, X = x] - \mathbb{E}[Y^0(0) \mid Z = 0, X = x] \nonumber \\
    &= \int y \left[f_{10}(y\mid x; \varphi) - f_{00}(y \mid x; \varphi)\right]\,\mathrm{d}y.
\end{align}
We assume $b(x; \varphi)$ is independent of $\varphi$ and thus estimate $b(x)$ and $\tau^h$ in different stages. This holds if mean exchangeability is assumed ($b(x;\varphi) = 0)$ or the true value of $b(x; \theta)$ is known.

Then we derive the corresponding score.
\begin{align}
    S(y, x, z, a; \varphi) =& S_f(x;\varphi) + \frac{z - e_Z(x; \varphi)}{e_Z(x;\varphi)[1 - e_Z(x;\varphi)]}\dot{e}_Z(x; \varphi) + \frac{z[a - e_A(x;\varphi)]}{e_A(x; \varphi)[1 - e_A(x;\varphi)]}\dot{e}_A(x;\varphi) \nonumber \\
    &+ (1 - z)S_{00}(y \mid x;\varphi) + (1-a)z S_{10}(y \mid x; \varphi) + az S_{11}(y \mid x; \varphi),
\end{align}
where $S_f(x;\varphi) = \frac{\partial \log p(x; \varphi)}{\partial \varphi}$, $S_{za}(y \mid x; \varphi) = \frac{\partial \log p_{za}(y \mid x; \varphi)}{\partial \varphi}$,  $\dot{e}_Z(x; \varphi) = \frac{\partial e_Z(x; \varphi)}{\partial \varphi}$ and $\dot{e}_A(x;\varphi) = \frac{\partial e_A(x;\varphi)}{\partial \varphi}$.
Then the tangent space is given by
\begin{align}
    \mathcal{T} = \Big\{& S(x) + (z-e_Z(x))g_Z(x) + z(a - e_A(x))g_A(x)  \nonumber\\
    &+ (1 - z)S_{00}(y \mid x) + (1-a)z S_{10}(y \mid x) + az S_{11}(y \mid x)\Big\},
\end{align}
where $\int S(x) p(x) \,\mathrm{d}x = 0$, $\int S_{za}(y\mid x)p_{za}(y \mid x)\,\mathrm{d} y = 0$, $\int y [S_{10}(y \mid x)p_{10}(y \mid x) - S_{00}(y \mid x)p_{00}(y \mid x)] = 0$, and $g_Z(x)$, $g_A(x)$ are arbitrary square-integrable measurable functions.

Under the parametric submodel $p(y, x, z, a;\varphi)$, $\tau^h$ can be calculated by 
\begin{align}
    \tau^h(\varphi) &= \frac{\mathbb{E}[h(e_Z(X))\{Y^1(1) - Y^1(0)\}]}{\mathbb{E}[h(e_Z(X))]} 
    \nonumber\\
    &= \frac{\iint y [f_{11}(y \mid x; \varphi) - f_{10}(y \mid x; \varphi)] h(e_Z(x;\varphi)) p(x; \varphi)\,\mathrm{d}y\mathrm{d}x }{\int h(e_Z(x;\varphi)) p(x;\varphi)\,\mathrm{d} x},
\end{align}
where $\tau^h = \tau^h(\varphi^*)$.
For notational simplicity, define $\Delta(x; \varphi) = \int y [p_{11}(y \mid x;\varphi) - p_{10}(y \mid x; \varphi)]\,\mathrm{d}y$, $u(\varphi) = \iint y [f_{11}(y \mid x; \varphi) - f_{10}(y \mid x; \varphi)] h(e_Z(x;\varphi)) p(x; \varphi)\,\mathrm{d}y\mathrm{d}x$, and $v(\varphi) = \int h(e_Z(x;\varphi)) p(x;\varphi)\,\mathrm{d} x$. Also, we omit $\varphi$ whenever the function is evaluated at $\varphi =\varphi^*$. Then, 
\begin{align}
    \frac{\partial \tau^h}{\partial \varphi} &= \frac{u'(\varphi) v(\varphi) - u(\varphi) v'(\varphi)}{v(\varphi)^2} = \frac{u'(\varphi) - \tau v'(\varphi)}{v(\varphi)},
\end{align}
and thus
\begin{align}
    \frac{\partial \tau^h}{\partial \varphi}\Big\vert_{\varphi=\varphi^*} = &\frac{\iint y[S_{11}(y\mid x)p_{11}(y\mid x)-S_{10}(y\mid x)p_{10}(y\mid x)] h(e_Z(x)) p(x)\,\mathrm{d}y\mathrm{d}x}{v} \noindent \\
    &+ \frac{\int \{\Delta(x) - \tau^h\}\dot{h}(e_Z(x))\dot{e}_Z(x)p(x)\,\mathrm{d}x}{v}
    + \frac{\int \{\Delta(x) - \tau^h\}h(e_Z(x))S(x)p(x)\,\mathrm{d}x}{v}.
\end{align}

We need to verify the following:
\begin{enumerate}
    \item $\mathrm{IF}(Y, X, Z, A) \in \mathcal{T}$;
    \item $\displaystyle\frac{\partial \tau^h}{\partial \varphi}\Big\vert_{\varphi=\varphi^*} = \mathbb{E}\left\{\mathrm{IF}(Y, X, Z, A)\cdot S(Y, X, Z, A; \varphi^*)\right\}$.
\end{enumerate}

To verify (1), note the following correspondence:
\begin{align}
    &S(X) = \frac{h(e_Z(X))}{\mathbb{E}[h(e_Z(X))]}(\Delta(X) - \tau^h); \\
    &g_Z(X) = \frac{\dot{h}(e_Z(X))}{\mathbb{E}[h(e_Z(X))]}(\Delta(X) - \tau^h); \\
    &g_A(X) = 0; \\
    &S_{11}(Y \mid X) = \frac{Y - \mu_{11}(X)}{e_Z(X)e_A(X)};\\
    &S_{10}(Y \mid X) = -\frac{Y - \mu_{10}(X)}{e_Z(X)[1 - e_A(X)] + [1 - e_Z(X)]r(X)};\\
    &S_{00}(Y \mid X) = -\frac{r(X)[Y - \mu_{00}(X)]}{e_Z(X)[1 - e_A(X)] + [1 - e_Z(X)]r(X)}.
\end{align}
Then, 
\begin{align*}
    \mathbb{E}[S_{za}(Y \mid X)\mid Z = z, A = a] &=  \mathbb{E}[\mathbb{E}[S_{za}(Y \mid X)\mid Z = z, A = a, X] \mid Z = z, A = a] = 0
\end{align*}
because $\mathbb{E}[Y \mid Z = z, A = a, X] = \mu_{za}(X)$. Also,
\begin{align*}
    \mathbb{E}[S(X)] = \frac{\mathbb{E}[h(e_Z(X))\Delta(x)]}{\mathbb{E}[h(e_Z(X))]} - \tau^h = 0.
\end{align*}

To verify (2), 
we have 
\begin{align}
    &\mathbb{E}\left\{\mathrm{IF}(Y, X, Z, A)\cdot S(Y, X, Z, A; \varphi^*)\right\} \\
    =&  \frac{1}{\mathbb{E}[h(e_Z(X))]}\Bigg\{\mathbb{E}\left[h(e_Z(X))\lambda(\Delta(X) - \tau^h)S(Y, X, Z, A)\right] \label{eq:1} \\
    &+ \mathbb{E}\left[h(e_Z(X)) \frac{ZA[Y - \mu_{11}(X)]S_{11}(Y \mid X)}{e_Z(X)e_A(X)}\right] \label{eq:2} \\
    &- \mathbb{E}\left[h(e_Z(X)) \frac{Z(1 - A)[Y - \mu_{10}(X)]S_{10}(Y \mid X)}{e_Z(X)[1 - e_A(X)] + [1 - e_Z(X)]r(X)}\right] \label{eq:3} \\
    &- \mathbb{E}\left[h(e_Z(X)) \frac{r(X)(1 - Z)[Y - \mu_{00}(X)]S_{00}(Y \mid X)}{e_Z(X)[1 - e_A(X)] + [1 - e_Z(X)]r(X)}\right]\Bigg\}. \label{eq:4}
\end{align}

Then, we calculate \eqref{eq:1} to \eqref{eq:4} respectively.
\begin{align}
    \eqref{eq:1} = &\mathbb{E}\Big[h(e_Z(X))\lambda(\Delta(X) - \tau^h)S(X)\Big] \label{eq:1-1}\\
    &+ \mathbb{E}\Big[h(e_Z(X))\lambda(\Delta(X) - \tau^h)\frac{Z - e_Z(X)}{e_Z(X)[1 - e_Z(X)]}\dot{e}_Z(X)\Big] \label{eq:1-2}\\
    &+ \mathbb{E}\Big[h(e_Z(X))\lambda(\Delta(X) - \tau^h)\frac{Z[A - e_A(X)]}{e_A(X)[1 - e_A(X)]}\dot{e}_A(X)\Big] \label{eq:1-3}\\
    &+ \mathbb{E}\Big[h(e_Z(X))\lambda(\Delta(X) - \tau^h)(1 - Z)S_{00}(Y \mid X)\Big] \label{eq:1-4}\\
    &+ \mathbb{E}\Big[h(e_Z(X))\lambda(\Delta(X) - \tau^h)(1 - A)ZS_{10}(Y \mid X)\Big] \label{eq:1-5}\\
    &+ \mathbb{E}\Big[h(e_Z(X))\lambda(\Delta(X) - \tau^h)AZS_{11}(Y \mid X)\Big].\label{eq:1-6}
\end{align}

We need to evaluate six terms from \eqref{eq:1-1} to \eqref{eq:1-6}.
\begin{align}
    \eqref{eq:1-1} = &  \mathbb{E}\Big[h(e_Z(X))(Z - e_Z(X))\frac{\dot{h}(e_Z(X))}{h(e_Z(X))}(\Delta(X) - \tau^h)S(X)\Big] \\
    &+ \mathbb{E}\Big[h(e_Z(X))(\Delta(X) - \tau^h)S(X)\Big] \\
    =& \mathbb{E}\Big[\dot{h}(e_Z(X))(\mathbb{E}[Z\mid X] - e_Z(X))(\Delta(X) - \tau^h)S(X)\Big] \\
    &+ \mathbb{E}\Big[h(e_Z(X))(\Delta(X) - \tau^h)S(X)\Big] \\
    =& \mathbb{E}\Big[h(e_Z(X))(\Delta(X) - \tau^h)S(X)\Big].
\end{align}
\begin{align}
    \eqref{eq:1-2} = &  \mathbb{E}\Big[h(e_Z(X))(Z - e_Z(X))\frac{\dot{h}(e_Z(X))}{h(e_Z(X))}(\Delta(X) - \tau^h)\frac{Z - e_Z(X)}{e_Z(X)[1 - e_Z(X)]}\dot{e}_Z(X)\Big] \\
    &+ \mathbb{E}\Big[h(e_Z(X))(\Delta(X) - \tau^h)\frac{Z - e_Z(X)}{e_Z(X)[1 - e_Z(X)]}\dot{e}_Z(X)\Big] \\
    =& \mathbb{E}\Big[\dot{h}(e_Z(X))(\Delta(X) - \tau^h)\frac{\mathbb{E}[(Z - e_Z(X))^2\mid X]}{e_Z(X)[1 - e_Z(X)]}\dot{e}_Z(X)\Big] \\
    &+ \mathbb{E}\Big[h(e_Z(X))(\Delta(X) - \tau^h)\frac{\mathbb{E}[Z - e_Z(X)\mid X]}{e_Z(X)[1 - e_Z(X)]}\dot{e}_Z(X)\Big] \\
    =& \mathbb{E}\Big[\dot{h}(e_Z(X))(\Delta(X) - \tau^h)\dot{e}_Z(X)\Big].
\end{align}
\begin{align}
    \eqref{eq:1-3} = &  \mathbb{E}\Big[h(e_Z(X))(Z - e_Z(X))\frac{\dot{h}(e_Z(X))}{h(e_Z(X))}(\Delta(X) - \tau^h)\frac{Z[A - e_A(X)]}{e_A(X)[1 - e_A(X)]}\dot{e}_A(X)\Big] \\
    &+ \mathbb{E}\Big[h(e_Z(X))(\Delta(X) - \tau^h)\frac{Z[A - e_A(X)]}{e_A(X)[1 - e_A(X)]}\dot{e}_A(X)\Big] \\
    =& \mathbb{E}\Big[\dot{h}(e_Z(X))(\Delta(X) - \tau^h)\frac{\mathbb{E}[Z(Z - e_Z(X))(A - e_A(X))\mid X]}{e_A(X)[1 - e_A(X)]}\dot{e}_A(X)\Big] \\
    &+ \mathbb{E}\Big[h(e_Z(X))(\Delta(X) - \tau^h)\frac{\mathbb{E}[Z(A - e_A(X))\mid X]}{e_A(X)[1 - e_A(X)]}\dot{e}_A(X)\Big] \\
    =& 0.
\end{align}
That $\eqref{eq:1-4} = \eqref{eq:1-5} = \eqref{eq:1-6}$ follows $\mathbb{E}[S_{za}(Y \mid X) \mid Z = z, A = a] = 0$. Hence,
\begin{equation}
    \eqref{eq:1} = \mathbb{E}\left\{[\Delta(X) - \tau^h]\left(h(e_Z(X))S(X) + \dot{h}(e_Z(X))\dot{e}_Z(X)\right)\right\}.
\end{equation}
\begin{align}
    \eqref{eq:2} &= \mathbb{E}\left[h(e_Z(X)) \frac{ZA[Y - \mu_{11}(X)]S_{11}(Y \mid X)}{e_Z(X)e_A(X)}\right] \\
    &= \mathbb{E}\left[h(e_Z(X)) \mathbb{E}[(Y - \mu_{11}(X))S_{11}(Y \mid X)\mid Z = 1, A = 1, X]\right] \\
    &= \mathbb{E}\left[h(e_Z(X)) \mathbb{E}[YS_{11}(Y \mid X)\mid Z = 1, A = 1, X]\right].
\end{align}
\begin{align}
    \eqref{eq:3} + \eqref{eq:4} &= -\mathbb{E}\left[h(e_Z(X)) \textstyle\frac{Z(1 - A)[Y - \mu_{10}(X)]S_{10}(Y \mid X) + r(X)(1 - Z)[Y - \mu_{00}(X)]S_{00}(Y\mid X)}{e_Z(X)[1 - e_A(X)] + [1 - e_Z(X)]r(X)}\right] \\
    &= -\mathbb{E}\left[h(e_Z(X)) \textstyle\frac{e_Z(X)[1 - e_A(X)]\mathbb{E}[YS_{10}(Y \mid X)\mid Z = 1, A = 0, X] + r(X)(1 - e_Z(X))\mathbb{E}[YS_{00}(Y\mid X)\mid Z = 0, X]}{e_Z(X)[1 - e_A(X)] + [1 - e_Z(X)]r(X)}\right] \\
    &= -\mathbb{E}\left[h(e_Z(X)) \mathbb{E}[YS_{10}(Y \mid X)\mid Z = 1, A = 0, X]\right].
\end{align}
Therefore, 
\begin{align}
    &\mathbb{E}\left\{\mathrm{IF}(Y, X, Z, A)\cdot S(Y, X, Z, A; \varphi^*)\right\} \\
    =&  \frac{1}{\mathbb{E}[h(e_Z(X))]}\Bigg\{\mathbb{E}\left\{[\Delta(X) - \tau^h]\left(h(e_Z(X))S(X) + \dot{h}(e_Z(X))\dot{e}_Z(X)\right)\right\} \\
    &+ \mathbb{E}\left[h(e_Z(X)) \mathbb{E}[YS_{11}(Y \mid X)\mid Z = 1, A = 1, X]\right]\\
    &- \mathbb{E}\left[h(e_Z(X)) \mathbb{E}[YS_{10}(Y \mid X)\mid Z = 1, A = 0, X]\right]\Bigg\} \\
    =&\frac{\partial \tau^h}{\partial \varphi}\Big\vert_{\varphi=\varphi^*}.
\end{align}

The EIF shown in \eqref{eq:EIF} simplifies to 
\begin{equation}
    \textstyle\mathrm{IF}(Y, X, Z, A) = \frac{h(e_Z(X))}{\mathbb{E}[h(e_Z(X)]}\Bigg\{
    \lambda(\Delta(X) - \tau^h) + \frac{ZA[Y - \mu_{11}(X)] }{e_Z(X) e_A(X)}
    - \frac{Z(1 - A)[Y - \mu_{10}(X)] + (1-Z)[Y - \mu_{00}(X)]}{1 - e_Z(X)e_A(X)}
    \Bigg\} \label{eq:EIF-simplified}
\end{equation}
if we assume $r(X) = 1$, i.e., $\mathrm{Var}[Y^1(X)\mid X] = \mathrm{Var}[Y^0(X) \mid X]$. Then the solution to $\sum_{i=1}^n \mathrm{IF}(Y_i, X_i, Z_i, A_i) = 0$ yields a consistent and locally efficient estimator
\begin{equation}
    \widehat{\tau}^h = \frac{\sum_{i=1}^n h(\widehat{e}_Z(X_i))\left[\widehat{\lambda}_i \widehat{\Delta}(X_i) + \widehat{T}(Y_i, X_i, Z_i, A_i)\right]}{\sum_{i=1}^n h(\widehat{e}_Z(X_i))\widehat{\lambda}_i},
\end{equation}
where 
\begin{align}
    \widehat{T}(Y_i, X_i, Z_i, A_i) &= \frac{Z_iA_i[Y_i - \widehat{\mu}_{11}(X_i)] }{\widehat{e}_Z(X_i) \widehat{e}_A(X_i)}
    - \frac{Z_i(1 - A_i)[Y_i - \widehat{\mu}_{10}(X_i)] + (1-Z_i)[Y_i - \widehat{\mu}_{00}(X_i)]}{1 - \widehat{e}_Z(X_i)\widehat{e}_A(X_i)}  \nonumber \\
    &=: \frac{\widehat{R}_{11i}}{\widehat{e}_{Zi}\widehat{e}_{Ai}} - \frac{\widehat{R}_{10i} + \widehat{R}_{00i}}{1 - \widehat{e}_{Zi}\widehat{e}_{Ai}}.
\end{align}

\section{Forms of $\widehat{\tau}^h$ for ATE and ATC in Section 5}

\paragraph{ATE}

The tilting function $h(e_Z(X)) = 1$, so
\begin{equation}
    \lambda = (Z - e_Z(X))\frac{\dot{h}(e_Z(X))}{h(e_Z(X))} + 1 = 1.
\end{equation}
Then, 
\begin{equation}
    \widehat{\tau}^{\ATE} = \frac{\sum_{i=1}^n (\widehat{\Delta}_i + \widehat{T}_i)}{n}.
\end{equation}

\paragraph{ATC}

The tilting function $h(e_Z(X)) = 1 - e_Z(X)$, so 
 \begin{equation}
    \lambda = -\frac{Z - e_Z(X)}{1 - e_Z(X)} + 1 = \frac{1 - Z}{1 - e_Z(X)}.
\end{equation}
Then, 
\begin{equation}
    \widehat{\tau}^{\ATC} = \frac{\sum_{i=1}^n [(1 - Z_i)\widehat{\Delta}_i + (1 - \widehat{e}_{Zi})\widehat{T}_i]}{\sum_{i=1}^n (1 - Z_i)}.
\end{equation}

\section{Proof of the unbiasedness of $\widehat{\tau}_\mathrm{ANCOVA}^\mathrm{const}$ under misspecification in Section 6.1}

The estimator $\widehat{\tau}_\mathrm{ANCOVA}^\mathrm{const}$ is the coefficient of $A$ in linear regression $Y \sim X + Z + A$, \emph{i.e}, the coefficient $\delta$ in $\mathbb{E}[Y \mid X, Z, A] = \alpha + X'\boldsymbol{\beta} + \gamma Z +\delta A$. Denote by $\boldsymbol{y}_{11}, \boldsymbol{y}_{10}, \boldsymbol{y}_{00}$ the observed outcomes for units in the concurrent treatment arm, concurrent control arm and the external control, respectively. Similarly, denote by $\mathbf{X}_{11}$, $\mathbf{X}_{10}$, $\mathbf{X}_{00}$ the covariate matrix for units in the concurrent treatment arm, concurrent control arm and the external control, respectively. Then, the the linear regression can be written as
\begin{equation}
    \begin{bmatrix}
        \boldsymbol{y}_{11} \\ \boldsymbol{y}_{10} \\ \boldsymbol{y}_{00}
    \end{bmatrix} = 
    \begin{bmatrix}
        \boldsymbol{1}_{11} & \mathbf{X}_{11} & \boldsymbol{1}_{11} & \boldsymbol{1}_{11} \\
        \boldsymbol{1}_{10} & \mathbf{X}_{10} & \boldsymbol{1}_{10} & \boldsymbol{0}_{10}\\
        \boldsymbol{1}_{00} & \mathbf{X}_{00} & \boldsymbol{0}_{00} & \boldsymbol{0}_{00}
    \end{bmatrix}
    \begin{bmatrix}
        \alpha \\ \boldsymbol{\beta} \\ \gamma \\ \delta
    \end{bmatrix} + \boldsymbol{\varepsilon},
\end{equation}
where $\boldsymbol{1}_{*}$ and $\boldsymbol{0}_{*}$ denote the all 1 and all 0 vector of size equal to the number of units in the $*$ group, respectively.
The normal equation shows that the OLS estimate should satisfy
\begin{equation}
    \begin{bmatrix}
        \boldsymbol{1}_{11}^\top & \boldsymbol{1}_{10}^\top & \boldsymbol{1}_{00}^\top \\
        \mathbf{X}_{11}^\top & \mathbf{X}_{10}^\top & \mathbf{X}_{00}^\top \\
        \boldsymbol{1}_{11}^\top & \boldsymbol{1}_{10}^\top & \boldsymbol{0}_{00}^\top \\
        \boldsymbol{1}_{11}^\top & \boldsymbol{0}_{10}^\top & \boldsymbol{0}_{00}^\top \\
    \end{bmatrix}
    \begin{bmatrix}
        \boldsymbol{1}_{11} & \mathbf{X}_{11} & \boldsymbol{1}_{11} & \boldsymbol{1}_{11} \\
        \boldsymbol{1}_{10} & \mathbf{X}_{10} & \boldsymbol{1}_{10} & \boldsymbol{0}_{10}\\
        \boldsymbol{1}_{00} & \mathbf{X}_{00} & \boldsymbol{0}_{00} & \boldsymbol{0}_{00}
    \end{bmatrix}
    \begin{bmatrix}
        \widehat{\alpha} \\ \widehat{\boldsymbol{\beta}} \\ \widehat{\gamma} \\ \widehat{\delta}
    \end{bmatrix} = \begin{bmatrix}
        \boldsymbol{1}_{11}^\top & \boldsymbol{1}_{10}^\top & \boldsymbol{1}_{00}^\top \\
        \mathbf{X}_{11}^\top & \mathbf{X}_{10}^\top & \mathbf{X}_{00}^\top \\
        \boldsymbol{1}_{11}^\top & \boldsymbol{1}_{10}^\top & \boldsymbol{0}_{00}^\top \\
        \boldsymbol{1}_{11}^\top & \boldsymbol{0}_{10}^\top & \boldsymbol{0}_{00}^\top \\
    \end{bmatrix}
    \begin{bmatrix}
        \boldsymbol{y}_{11} \\ \boldsymbol{y}_{10} \\ \boldsymbol{y}_{00}
    \end{bmatrix},
\end{equation}
which yields
\begin{equation}
    \begin{bmatrix}
        N & \mathbf{1}^\top \mathbf{X} & N_{11} + N_{10} & N_{11} \\
        \mathbf{X}^\top \mathbf{1} & \mathbf{X}^\top \mathbf{X} & \mathbf{X}_{11}^\top \mathbf{1}_{11} + \mathbf{X}_{10}^\top \mathbf{1}_{10} & \mathbf{X}_{11}^\top \mathbf{1}_{11} \\
        N_{11} + N_{10} & \mathbf{1}_{11}^\top \mathbf{X}_{11} + \mathbf{1}_{10}^\top \mathbf{X}_{10} & N_{11} + N_{10} & N_{11} \\
        N_{11} & \mathbf{1}_{11}^\top \mathbf{X}_{11} & N_{11} & N_{11}
    \end{bmatrix} \begin{bmatrix}
        \widehat{\alpha} \\ \widehat{\boldsymbol{\beta}} \\ \widehat{\gamma} \\ \widehat{\delta}
    \end{bmatrix} = 
     \begin{bmatrix}
        \boldsymbol{1}^\top \boldsymbol{y} \\ \mathbf{X}^\top \boldsymbol{y} \\
        \boldsymbol{1}_{11}^\top \boldsymbol{y}_{11} + \boldsymbol{1}_{10}^\top \boldsymbol{y}_{10} \\ \boldsymbol{1}_{11}^\top \boldsymbol{y}_{11}
    \end{bmatrix}
\end{equation}

We subtract the third row from the first row, and subtract the fourth row from the third row. By rearranging the rows, we have
\begin{equation}
    \begin{bmatrix}
        \mathbf{X}^\top \mathbf{1} & \mathbf{X}^\top \mathbf{X} & \mathbf{X}_{11}^\top \mathbf{1}_{11} + \mathbf{X}_{10}^\top \mathbf{1}_{10} & \mathbf{X}_{11}^\top \mathbf{1}_{11} \\
        N_{00} & \mathbf{1}_{00}^\top \mathbf{X}_{00} & N_{00} & 0 \\
        N_{10} & \mathbf{1}_{10}^\top \mathbf{X}_{10} & N_{10} & 0 \\
        N_{11} & \mathbf{1}_{11}^\top \mathbf{X}_{11} & N_{11} & N_{11}
    \end{bmatrix} \begin{bmatrix}
        \widehat{\alpha} \\ \widehat{\boldsymbol{\beta}} \\ \widehat{\gamma} \\ \widehat{\delta}
    \end{bmatrix} = 
     \begin{bmatrix}
        \mathbf{X}^\top \boldsymbol{y} \\
        \boldsymbol{1}_{00}^\top \boldsymbol{y}_{00} \\ 
        \boldsymbol{1}_{10}^\top \boldsymbol{y}_{10} \\ \boldsymbol{1}_{11}^\top \boldsymbol{y}_{11}
    \end{bmatrix}
\end{equation}
The last two rows state that 
\begin{align}
    \widehat{\alpha} + \frac{\boldsymbol{1}_{10}^\top \mathbf{X}_{10}}{N_{10}} \widehat{\boldsymbol{\beta}} + \widehat{\gamma} =  \frac{\boldsymbol{1}_{10}^\top \boldsymbol{y}_{10}}{N_{10}}, \quad 
    \widehat{\alpha} + \frac{\boldsymbol{1}_{11}^\top \mathbf{X}_{11}}{N_{11}} \widehat{\boldsymbol{\beta}} + \widehat{\gamma} + \widehat{\delta} = \frac{\boldsymbol{1}_{11}^\top \boldsymbol{y}_{11}}{N_{11}}.
\end{align}
Therefore, 
\begin{equation}
    \widehat{\delta} = \frac{\boldsymbol{1}_{11}^\top \boldsymbol{y}_{11}}{N_{11}} - \frac{\boldsymbol{1}_{10}^\top \boldsymbol{y}_{10}}{N_{11}} - \frac{\boldsymbol{1}_{11}^\top \mathbf{X}_{11}}{N_{11}} \widehat{\boldsymbol{\beta}} + \frac{\boldsymbol{1}_{10}^\top \mathbf{X}_{10}}{N_{10}} \widehat{\boldsymbol{\beta}} = \overline{\boldsymbol{y}}_{11} - \overline{\boldsymbol{y}}_{10} - \left(\overline{\mathbf{X}}_{11} - \overline{\mathbf{X}}_{10}\right)\widehat{\boldsymbol{\beta}}.
\end{equation}
Since the concurrent study is an RCT, the covariates in the treatment and control arm should be identically distributed. Hence, $\mathbb{E}[\overline{\mathbf{X}}_{11} - \overline{\mathbf{X}}_{10}] = 0$ and 
\begin{equation}
    \mathbb{E}[\widehat{\delta}] = \mathbb{E}[\overline{\boldsymbol{y}}_{11} - \overline{\boldsymbol{y}}_{10}] = \tau^{\ATT}.
\end{equation}

\section{Additional tables}

\paragraph{Simulation 2 with stronger covariate predictability} We replicate Simulation 2 with the standard deviation of the error term for $Y$ being 0.2 instead of 1.

\begin{table}[h]
    \centering
    \resizebox{\textwidth}{!}{
        \begin{tabular}{cccccccccccccc}
            \toprule
            & & \multicolumn{2}{c}{RCT only} & \multicolumn{7}{c}{With external control}\\
            \cmidrule(lr){3-4} \cmidrule(lr){5-11}
            $b$ & 1:$m$ & $\widehat{\tau}_{\mathrm{MD}}$ & $\widehat{\tau}_{\mathrm{MDP}}$ & $\widehat{\tau}_{\mathrm{PS}}$ & $\widehat{\tau}_{\mathrm{DR}}$ & $\widehat{\tau}_{\mathrm{ANCOVA}}^{\mathrm{ME}}$ & $\widehat{\tau}_{\mathrm{ANCOVA}}^{\mathrm{const}}$ & $\widehat{\tau}_{aug}^{\mathrm{ME}}$ & $\widehat{\tau}_{aug}^{\mathrm{const}}$ & $\widehat{\tau}_{aug}^{\mathrm{flex}}$\\ 
            \midrule
            \multirow{5}{*}{0} & 1:1 & 0 (10) & 0 (1) & -1 (9) & 0 (1) & 1 (3) & 0 (2) & 0 (1) & 0 (1) & 0 (1) \\
             & 1:2 & 0 (10) & 0 (1) & 0 (8) & 0 (1) & 1 (2) & 0 (1) & 0 (1) & 0 (1) & 0 (1) \\
             & 1:5 & 0 (12) & 0 (2) & -1 (8) & 0 (2) & 1 (3) & 0 (3) & 0 (1) & 0 (2) & 0 (2) \\
             & 1:10 & 0 (16) & 0 (3) & -1 (8) & 0 (3) & 2 (3) & 0 (6) & 0 (2) & 0 (3) & 0 (3) \\
             & 1:20 & 0 (21) & 0 (5) & -1 (8) & 0 (5) & 2 (3) & 0 (9) & 0 (2) & 0 (4) & 0 (5) \\
            \midrule
            \multirow{5}{*}{20} & 1:1 & 0 (10) & 0 (1) & 17 (11) & 0 (1) & 11 (4) & 0 (3) & 11 (2) & 0 (2) & 0 (1) \\
             & 1:2 & 0 (10) & 0 (1) & 20 (10) & 0 (2) & 14 (3) & 0 (3) & 13 (2) & -1 (3) & 0 (1) \\
             & 1:5 & 0 (12) & 0 (2) & 23 (10) & 0 (2) & 17 (3) & 0 (5) & 17 (3) & -1 (5) & 0 (2) \\
             & 1:10 & 0 (16) & 0 (3) & 24 (11) & 0 (4) & 19 (4) & 0 (8) & 20 (3) & -1 (8) & 0 (3) \\
             & 1:20 & 0 (21) & 0 (5) & 25 (11) & 0 (7) & 20 (4) & 0 (12) & 22 (4) & -1 (12) & 0 (5) \\
            \midrule
            \multirow{5}{*}{40} & 1:1 & 0 (10) & 0 (1) & 35 (13) & 0 (1) & 22 (5) & 0 (5) & 22 (3) & -1 (3) & 0 (1) \\
             & 1:2 & 0 (10) & 0 (1) & 40 (13) & 0 (2) & 26 (5) & 0 (6) & 26 (4) & -1 (5) & 0 (1) \\
             & 1:5 & 0 (12) & 0 (2) & 46 (14) & 0 (3) & 33 (5) & 0 (8) & 34 (5) & -2 (9) & 0 (2) \\
             & 1:10 & 0 (16) & 0 (3) & 49 (15) & 0 (6) & 36 (5) & 0 (12) & 40 (6) & -2 (15) & 0 (3) \\
             & 1:20 & 0 (21) & 0 (5) & 51 (15) & 1 (12) & 39 (5) & 0 (17) & 45 (6) & -2 (23) & 0 (5) \\
            \bottomrule\\
        \end{tabular}
    }
    \caption{Bias and standard deviation (in brackets) of the estimators described in Section 6.2 under different values of the systematic difference $b$ and treatment-to-control ratio $m$ in Simulation 2 under heterogeneous treatment effects. Here, the standard deviation of the error term on $Y$ is reduced to 0.2 from 1. The estimators include two estimators that only use the RCT data: the mean difference $\widehat{\tau}_{\mathrm{MD}}$ and the mean difference in model-based predicted outcome $\widehat{\tau}_{\mathrm{MDP}}$, and seven estimators that utilize external control, including the propensity score weighted estimator $\widehat{\tau}_{\mathrm{PS}}$, the doubly robust estimator $\widehat{\tau}_{\mathrm{DR}}$, the ANCOVA estimators $\widehat{\tau}_{\mathrm{ANCOVA}}^{\mathrm{ME}}$ and $\widehat{\tau}_{\mathrm{ANCOVA}}^{\mathrm{const}}$ without and with a separate intercept for $Z$ respectively,
    and the proposed augmented estimators, $\widehat{\tau}_{\mathrm{ATT}}^{\mathrm{ME}}$, $\widehat{\tau}_{\mathrm{ATT}}^{\mathrm{const}}$ and $\widehat{\tau}_{\mathrm{ATT}}^{\mathrm{flex}}$ with increasing flexibility of $b(x)$. All columns except 1:$m$ are multiplied by 100 and rounded to the nearest integer.}
    \label{tab:sim_2_supplement}
\end{table}

\paragraph{Simulation 3}

\begin{table}[h]
    \centering
    \resizebox{\textwidth}{!}{
        \begin{tabular}{cccccccccccccc}
            \toprule
            & & \multicolumn{2}{c}{RCT only} & \multicolumn{7}{c}{With external control}\\
            \cmidrule(lr){3-4} \cmidrule(lr){5-11}
            $b$ & 1:$m$ & $\widehat{\tau}_{\mathrm{MD}}$ & $\widehat{\tau}_{\mathrm{MDP}}$ & $\widehat{\tau}_{\mathrm{PS}}$ & $\widehat{\tau}_{\mathrm{DR}}$ & $\widehat{\tau}_{\mathrm{ANCOVA}}^{\mathrm{ME}}$ & $\widehat{\tau}_{\mathrm{ANCOVA}}^{\mathrm{const}}$ & $\widehat{\tau}_{aug}^{\mathrm{ME}}$ & $\widehat{\tau}_{aug}^{\mathrm{const}}$ & $\widehat{\tau}_{aug}^{\mathrm{flex}}$\\ 
            \midrule
            \multirow{5}{*}{0} & 1:1 & -1 (16) & 0 (6) & -1 (16) & 0 (11) & 1 (10) & 0 (12) & 0 (10) & -1 (11) & -1 (11) \\
             & 1:2 & 0 (17) & 0 (7) & 0 (17) & 0 (13) & 2 (11) & 0 (12) & 0 (11) & -1 (13) & 0 (13) \\
             & 1:5 & 0 (22) & 0 (10) & 0 (17) & 1 (18) & 3 (11) & 0 (18) & 1 (12) & -1 (18) & 1 (18) \\
             & 1:10 & 1 (30) & 0 (16) & 0 (18) & 3 (26) & 3 (11) & 1 (25) & 1 (13) & -1 (25) & 3 (26) \\
             & 1:20 & 0 (40) & -1 (26) & 0 (18) & 4 (37) & 3 (12) & 0 (34) & 1 (14) & -3 (34) & 4 (39) \\
            \midrule
            \multirow{5}{*}{20} & 1:1 & -1 (16) & 0 (6) & 18 (19) & 0 (11) & 12 (10) & 0 (11) & 11 (10) & -2 (11) & -1 (11) \\
             & 1:2 & 0 (17) & 0 (7) & 20 (20) & 0 (13) & 15 (11) & 0 (12) & 14 (11) & -2 (13) & 0 (13) \\
             & 1:5 & 0 (22) & 0 (10) & 23 (21) & 1 (18) & 19 (12) & 0 (18) & 18 (13) & -3 (18) & 1 (18) \\
             & 1:10 & 1 (30) & 0 (16) & 25 (22) & 3 (26) & 21 (12) & 1 (25) & 21 (14) & -3 (25) & 2 (26) \\
             & 1:20 & 0 (40) & -1 (26) & 26 (23) & 4 (37) & 22 (13) & 0 (35) & 23 (16) & -4 (34) & 4 (39) \\
            \midrule
            \multirow{5}{*}{40} & 1:1 & -1 (16) & 0 (6) & 36 (22) & 0 (11) & 23 (11) & 0 (12) & 22 (11) & -2 (12) & -1 (11) \\
             & 1:2 & 0 (17) & 0 (7) & 41 (23) & 0 (13) & 28 (12) & 0 (13) & 27 (12) & -3 (13) & -1 (13) \\
             & 1:5 & 0 (22) & 0 (10) & 47 (25) & 1 (18) & 35 (13) & 1 (18) & 35 (14) & -4 (19) & 0 (18) \\
             & 1:10 & 1 (30) & 0 (16) & 50 (26) & 3 (26) & 39 (14) & 1 (26) & 41 (16) & -4 (27) & 2 (27) \\
             & 1:20 & 0 (40) & -1 (26) & 52 (27) & 5 (38) & 41 (14) & 0 (36) & 46 (18) & -6 (38) & 3 (39) \\
            \bottomrule\\
        \end{tabular}
    }
    \caption{Bias and standard deviation (in brackets) of the estimators described in Section 6.3 under different values of the systematic difference $b$ and treatment-to-control ratio $m$ in Simulation 3 under heterogeneous treatment effects. The estimators include two estimators that only use the RCT data: the mean difference $\widehat{\tau}_{\mathrm{MD}}$ and the mean difference in model-based predicted outcome $\widehat{\tau}_{\mathrm{MDP}}$, and seven estimators that utilize external control, including the propensity score weighted estimator $\widehat{\tau}_{\mathrm{PS}}$, the doubly robust estimator $\widehat{\tau}_{\mathrm{DR}}$, the ANCOVA estimators $\widehat{\tau}_{\mathrm{ANCOVA}}^{\mathrm{ME}}$ and $\widehat{\tau}_{\mathrm{ANCOVA}}^{\mathrm{const}}$ without and with a separate intercept for $Z$ respectively,
    and the proposed augmented estimators, $\widehat{\tau}_{\mathrm{ATT}}^{\mathrm{ME}}$, $\widehat{\tau}_{\mathrm{ATT}}^{\mathrm{const}}$ and $\widehat{\tau}_{\mathrm{ATT}}^{\mathrm{flex}}$ with increasing flexibility of $b(x)$. All columns except 1:$m$ are multiplied by 100 and rounded to the nearest integer.}
\end{table}

\paragraph{Estimands ATE, ATC and ATO with same $b(X)$ as in Simulation 2}

\begin{table}[h]
    \centering
    \resizebox{\textwidth}{!}{
        \begin{tabular}{cccccccccccccc}
            \toprule
            $b$ & 1:$m$ & $\widehat{\tau}_{\mathrm{ATE}}^{\mathrm{ME}}$ & $\widehat{\tau}_{\mathrm{ATE}}^{\mathrm{const}}$ & $\widehat{\tau}_{\mathrm{ATE}}^{\mathrm{flex}}$ & $\widehat{\tau}_{\mathrm{ATC}}^{\mathrm{ME}}$ & $\widehat{\tau}_{\mathrm{ATC}}^{\mathrm{const}}$ & $\widehat{\tau}_{\mathrm{ATC}}^{\mathrm{flex}}$ & $\widehat{\tau}_{\mathrm{ATO}}^{\mathrm{ME}}$ & $\widehat{\tau}_{\mathrm{ATO}}^{\mathrm{const}}$ & $\widehat{\tau}_{\mathrm{ATO}}^{\mathrm{flex}}$\\ 
            \midrule
            \multirow{5}{*}{0} & 1:1 & 0 (9) & -1 (10) & -1 (11) & -1 (16) & -1 (18) & -1 (19) & 0 (8) & -1 (9) & -1 (10) \\
             & 1:2 & 0 (8) & 0 (10) & 0 (11) & 0 (14) & 0 (17) & 0 (18) & 0 (8) & 0 (10) & 0 (10) \\
             & 1:5 & 0 (8) & 0 (12) & 0 (14) & 0 (13) & 0 (17) & 0 (21) & 0 (8) & 0 (12) & 0 (13) \\
             & 1:10 & 0 (8) & 0 (17) & 0 (19) & 0 (13) & 0 (21) & 0 (27) & 0 (8) & 0 (16) & 0 (19) \\
             & 1:20 & 0 (8) & -1 (23) & -1 (29) & 0 (13) & -1 (25) & -1 (40) & 0 (8) & -1 (22) & -1 (28) \\
            \midrule
            \multirow{5}{*}{20} & 1:1 & 9 (10) & -6 (11) & -1 (11) & 6 (16) & -11 (18) & -1 (19) & 10 (9) & -4 (9) & -1 (10) \\
             & 1:2 & 11 (8) & -6 (11) & 0 (11) & 8 (15) & -11 (17) & 0 (18) & 12 (8) & -4 (10) & 0 (10) \\
             & 1:5 & 13 (8) & -7 (13) & 0 (14) & 10 (13) & -12 (18) & 0 (21) & 15 (8) & -6 (13) & 0 (13) \\
             & 1:10 & 15 (8) & -7 (18) & 0 (19) & 11 (13) & -13 (22) & 0 (27) & 17 (8) & -6 (18) & 0 (19) \\
             & 1:20 & 17 (8) & -8 (25) & -1 (29) & 11 (13) & -15 (28) & -1 (40) & 18 (8) & -8 (25) & -1 (28) \\
            \midrule
            \multirow{5}{*}{40} & 1:1 & 18 (10) & -11 (11) & -1 (11) & 14 (17) & -20 (19) & -1 (19) & 21 (9) & -8 (10) & -1 (10) \\
             & 1:2 & 21 (9) & -12 (12) & 0 (11) & 17 (15) & -22 (19) & 0 (18) & 25 (9) & -9 (12) & 0 (10) \\
             & 1:5 & 27 (9) & -13 (16) & 0 (14) & 20 (14) & -25 (21) & 0 (21) & 30 (9) & -11 (16) & 0 (13) \\
             & 1:10 & 31 (9) & -14 (23) & 0 (19) & 22 (14) & -27 (27) & 0 (27) & 33 (9) & -13 (23) & 0 (19) \\
             & 1:20 & 34 (9) & -16 (31) & -1 (29) & 23 (14) & -29 (34) & -1 (40) & 36 (9) & -14 (32) & -1 (28) \\
            \bottomrule\\
        \end{tabular}
    }
    \caption{Bias and standard deviation (in brackets) of the estimators described in Section 6.4 under different values of the systematic difference $b$ and treatment-to-control ratio $m$ in Simulation 4 under heterogeneous treatment effects. Here, the subscripts $\mathrm{ATE}, \mathrm{ATC}$ and $\mathrm{ATO}$ refer to the estimands, and the superscripts $\mathrm{ME}$, $\mathrm{const}$ and $\mathrm{flex}$ correspond to the proposed method assuming $b(X) = 0$, $b(X)$ is constant and $b(X)$ does not have a particular form, respectively. All columns except 1:$m$ are multiplied by 100 and rounded to the nearest integer.}
    \label{tab:sim_4}
\end{table}

\paragraph{Estimands ATE, ATC and ATO with constant $b(X)$}

We replicate Simulation 4, but with $\beta_{00} = \beta_{10} = (-0.6, 0.8, -0.9, -0.7)'$ and $\beta_{11} = (-0.8, 0.1, -0.5, -1.1)$. Here $b(X) = b$ is a constant.

\begin{table}[h]
    \centering
    \resizebox{\textwidth}{!}{
        \begin{tabular}{cccccccccccccc}
            \toprule
            $b$ & 1:$m$ & $\widehat{\tau}_{\mathrm{ATE}}^{\mathrm{ME}}$ & $\widehat{\tau}_{\mathrm{ATE}}^{\mathrm{const}}$ & $\widehat{\tau}_{\mathrm{ATE}}^{\mathrm{flex}}$ & $\widehat{\tau}_{\mathrm{ATC}}^{\mathrm{ME}}$ & $\widehat{\tau}_{\mathrm{ATC}}^{\mathrm{const}}$ & $\widehat{\tau}_{\mathrm{ATC}}^{\mathrm{flex}}$ & $\widehat{\tau}_{\mathrm{ATO}}^{\mathrm{ME}}$ & $\widehat{\tau}_{\mathrm{ATO}}^{\mathrm{const}}$ & $\widehat{\tau}_{\mathrm{ATO}}^{\mathrm{flex}}$\\ 
            \midrule
            \multirow{5}{*}{0} & 1:1 & 0 (9) & -1 (10) & -1 (11) & -1 (16) & -1 (18) & -1 (19) & 0 (9) & -1 (10) & -1 (10) \\
             & 1:2 & 0 (8) & 0 (10) & 0 (11) & 0 (14) & 0 (17) & 0 (18) & 0 (8) & 0 (10) & 0 (10) \\
             & 1:5 & 0 (8) & 0 (12) & 0 (14) & 0 (13) & 0 (17) & 0 (21) & 0 (8) & 0 (12) & 0 (13) \\
             & 1:10 & 0 (8) & 0 (17) & 0 (19) & 0 (13) & 0 (21) & 0 (27) & 0 (8) & 0 (16) & 0 (19) \\
             & 1:20 & 0 (8) & -1 (23) & -1 (29) & 0 (13) & -1 (25) & -1 (40) & 0 (8) & -1 (22) & -1 (28) \\
            \midrule
            \multirow{5}{*}{20} & 1:1 & 12 (9) & -1 (10) & -1 (11) & 14 (16) & -1 (18) & -1 (19) & 12 (9) & -1 (10) & -1 (10) \\
             & 1:2 & 14 (8) & 0 (10) & 0 (11) & 16 (14) & 0 (17) & 0 (18) & 14 (8) & 0 (10) & 0 (10) \\
             & 1:5 & 16 (8) & 0 (12) & 0 (14) & 18 (13) & 0 (17) & 0 (21) & 16 (8) & 0 (12) & 0 (13) \\
             & 1:10 & 17 (8) & 0 (17) & 0 (19) & 19 (13) & 0 (21) & 0 (27) & 18 (8) & 0 (16) & 0 (19) \\
             & 1:20 & 18 (8) & -1 (23) & -1 (29) & 19 (13) & -1 (25) & -1 (40) & 18 (8) & -1 (22) & -1 (28) \\
            \midrule
            \multirow{5}{*}{40} & 1:1 & 25 (9) & -1 (10) & -1 (11) & 29 (16) & -1 (18) & -1 (19) & 25 (9) & -1 (10) & -1 (10) \\
             & 1:2 & 28 (8) & 0 (10) & 0 (11) & 32 (14) & 0 (17) & 0 (18) & 28 (8) & 0 (10) & 0 (10) \\
             & 1:5 & 32 (8) & 0 (12) & 0 (14) & 35 (13) & 0 (17) & 0 (21) & 33 (8) & 0 (12) & 0 (13) \\
             & 1:10 & 34 (8) & 0 (17) & 0 (19) & 37 (13) & 0 (21) & 0 (27) & 35 (8) & 0 (16) & 0 (19) \\
             & 1:20 & 36 (8) & -1 (23) & -1 (29) & 38 (13) & -1 (25) & -1 (40) & 37 (8) & -1 (22) & -1 (28) \\
            \bottomrule\\
        \end{tabular}
    }
     \caption{Bias and standard deviation (in brackets) of the estimators described in Section 6.4 under different values of the systematic difference $b$ and treatment-to-control ratio $m$ in Simulation 4 under heterogeneous treatment effects, but with $b(X) = b$. Here, the subscripts $\mathrm{ATE}, \mathrm{ATC}$ and $\mathrm{ATO}$ refer to the estimands, and the superscripts $\mathrm{ME}$, $\mathrm{const}$ and $\mathrm{flex}$ correspond to the proposed method assuming $b(X) = 0$, $b(X)$ is constant and $b(X)$ does not have a particular form, respectively. All columns except 1:$m$ are multiplied by 100 and rounded to the nearest integer.}
    \label{tab:sim_4_supplement}
\end{table}
